\documentclass[11pt,twocolumn]{article}
\usepackage[margin=0.75in]{geometry}
\usepackage{amsmath,amssymb}
\usepackage{booktabs}
\usepackage{graphicx}
\usepackage{siunitx}
\usepackage{hyperref}
\usepackage{natbib}
\usepackage{xcolor}
\usepackage{float}
\usepackage{multirow}
\usepackage{cuted}
\usepackage{flushend}
\usepackage[capitalise]{cleveref}
\crefname{subsection}{Section}{Sections}
\title{Energy-efficient Codon Optimization on Thermodynamic Hardware}

\author{
Andraž Jelinčič,$^{1,}$\footnote{Corresponding author: \texttt{andraz@extropic.ai}} \hspace{5mm} Ross C.\ Walker$^{1}$ \\[2mm]
\small $^{1}$Extropic Corp., Boston, MA \\
\small \texttt{andraz@extropic.ai},
\texttt{rwalker@extropic.ai}
}

\date{}

\begin{document}

\twocolumn[
  \maketitle
  \begin{@twocolumnfalse}
  \begin{abstract}
  \noindent
  The growing energy demand for computation is becoming increasingly unsustainable.
  Thermodynamic computing, which harnesses physical thermal fluctuations as a computational resource rather than suppressing them, offers orders-of-magnitude energy savings for probabilistic and combinatorial tasks.
  Pharmaceutical R\&D, heavily reliant on computational optimization and sampling, is a natural application domain.
  Here we present what is, to our knowledge, the first concrete pharmaceutical application mapped to thermodynamic hardware with energy estimates grounded in prototype measurements.
  We reduce mRNA codon optimization, a combinatorial problem routinely solved in drug development, to sampling from an Ising model, making it directly executable on a thermodynamic sampling unit (TSU).
  Benchmarking three approaches (Potts sampling, Ising sampling, and a genetic algorithm baseline) on the SARS-CoV-2 spike protein, we find that all achieve comparable optimization quality (scores ${\sim}234$--$240$), but energy estimates based on validated hardware models indicate that a TSU could solve this problem using approximately $10^6\times$ less energy than a conventional GPU. 
  All code is released under an open-source license.
  \end{abstract}
  \vspace{1em}
  \end{@twocolumnfalse}
]

\section{Introduction}

Recent large-scale investments in artificial intelligence and computational workloads are placing increasing strain on the global energy infrastructure.
Every year, U.S.\ companies spend an amount exceeding the inflation-adjusted cost of the Apollo program on AI-focused data centers. With demand projected to grow from 59~GW in 2025 to more than 120~GW by 2030~\cite{iea2024}, there are concerns that the energy demand for data centers is becoming unsustainable.
At the same time, demand driven increases in model sizes and deployment footprints continue to grow rapidly, creating an urgent need for more efficient computing paradigms.

Existing AI systems are optimized for GPU hardware which was originally designed for computer graphics and whose suitability for machine learning was only discovered accidentally decades later.
Had different cost effective hardware architectures been available, algorithms may well have evolved in a different and possibly more energy-efficient direction.
This interplay between algorithm research and hardware availability, known as the ``hardware lottery''~\cite{hooker2021hardware}, entrenches hardware-algorithm pairings that may be far from optimal.
Prudent planning calls for systematic exploration of alternative computing architectures~\cite{jelincic2025dtm}.

\begin{figure*}[tb]
\centering
\includegraphics[width=\textwidth]{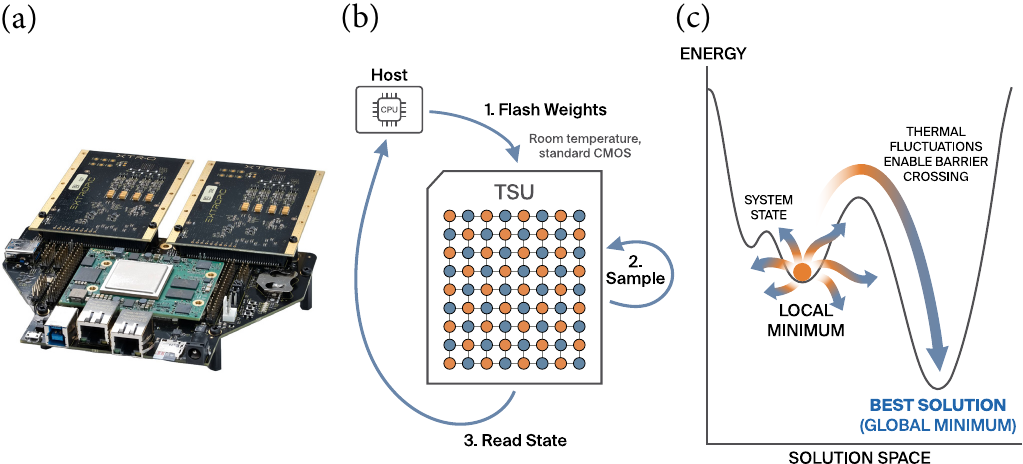}

\caption{\textbf{Thermodynamic computing overview.}
\textbf{(a)}~A picture of a system containing two of Extropic's prototype thermodynamic computing chips. These chips implement probabilistic computing primitives and are used to confirm the correct operation of these primitives and to measure their energy consumption.
\textbf{(b)}~Schematic of a thermodynamic sampling unit (TSU). The chip contains a grid of interconnected probabilistic bits (p-bits), two-colored for parallel block Gibbs updates. The host programs weights, the chip performs rapid sampling, and the host reads back the result.
\textbf{(c)}~Stochastic optimization on an energy landscape. Thermal fluctuations enable the system to explore broadly and escape local minima, converging toward low-energy solutions.}
\label{fig:thermo-overview}
\end{figure*}

Thermodynamic computing offers one such alternative~\cite{hylton2019thermo, hylton2020thermo}. In nature, thermodynamic laws govern the behavior of large collections of interacting atoms, which inevitably tend to settle into stable arrangements with near-minimum energy. Many famous optimization algorithms including Langevin Monte Carlo, Simulated Annealing, and Diffusion Modeling, are inspired by (and in some cases, exactly simulating) this phenomenon. The goal of thermodynamic computing hardware is to target and accelerate algorithms of this type.

More specifically, this hardware can be used to perform an \textit{analog simulation} of the physical dynamics that would arise in a system governed by a user-defined energy function.
Allowing this simulation to relax toward thermal equilibrium produces samples from the corresponding \textit{Boltzmann distribution}. This stands in sharp contrast to conventional computing, where
GPUs and CPUs expend substantial energy suppressing thermal noise, an intrinsic property of all integrated circuits, in order to guarantee deterministic operation.
When such hardware is then used to run inherently probabilistic algorithms, the result is a fundamental inefficiency: considerable energy is spent enforcing perfect determinism, only for the software to immediately re-inject randomness through pseudo-random number generation.
Thermodynamic hardware circumvents this logic entirely, harnessing the very thermal fluctuations that conventional processors expend energy to suppress, and thereby performing natively what conventional hardware must laboriously simulate~\cite{jelincic2025dtm}.

The physical embodiment of this approach is the thermodynamic sampling unit (TSU): a programmable chip fabricated in standard CMOS, operating at room temperature, that contains arrays of probabilistic computing elements connected in a sparse, locally-connected graph (\cref{fig:thermo-overview}).
A TSU is built from a family of primitive sampling circuits (\cref{tab:tsu-primitives}), each of which natively samples from a different distribution class: p-bits (Bernoulli), p-dits (categorical), p-modes (Gaussian), and p-MoGs (mixture of Gaussians).
By programming the interaction weights between these elements, the chip's thermal fluctuations explore the corresponding Boltzmann distribution.
This work uses p-bits for the Ising formulation and, prospectively, p-dits for the native Potts formulation.
A TSU operates in a flash-sample-read cycle: a host computer programs the interaction weights, the chip performs many rounds of Gibbs sampling at extreme speed and low energy, and the host reads back the resulting state.

\begin{table*}[h]
\centering
\caption{Probabilistic primitives of a TSU.}
\label{tab:tsu-primitives}
\vspace{2mm}
\begin{tabular}{lll}
\toprule
Primitive & Distribution & Hardware implementation \\
\midrule
p-bit & Bernoulli (binary) & Subthreshold transistor RNG with sigmoidal bias \\
p-dit & Categorical ($K$-state) & Softmax sampling circuit \\
p-mode & Gaussian (continuous) & Analog noise source with programmable mean/variance \\
p-MoG & Mixture of Gaussians & Array of coupled p-modes \\
\bottomrule
\end{tabular}
\end{table*}
Recent work has demonstrated that a device based on this architecture could achieve performance parity with GPUs on a generative image benchmark while using approximately $10{,}000\times$ less energy per sample~\cite{jelincic2025dtm}. Real world measurements from prototype hardware have validated the underlying energy models.
Prototype hardware is already in production and a TSU capable of running the algorithms described in this paper is projected to be commercially available within two years.

\paragraph{Pharmaceutical applications.}
Pharmaceutical research and development is a natural application domain for thermodynamic computing.
Drug discovery relies heavily on computational optimization and sampling. For example: conformational search, binding affinity estimation, molecular design, protein folding and sequence optimization all reduce to energy minimization or probabilistic inference over discrete or continuous variables. These are precisely the problems for which thermodynamic hardware is designed.
Conventional hardware such as GPUs, optimized for deterministic matrix multiplication, is inherently inefficient for these natively probabilistic workloads. Nevertheless, to our knowledge, no prior work has demonstrated the potential applications of thermodynamic computing to discrete optimization problems in drug design nor provided energy estimates grounded in real measurements.

In this work, we provide such a demonstration using mRNA codon optimization as an example real world workload.
In all living organisms, proteins are synthesized by ribosomes that read messenger RNA (mRNA) in triplets called \emph{codons}, each coding for one amino acid. Because the genetic code is degenerate, with 61 codons encoding for only 20 amino acids, most amino acids can be specified by two to six \emph{synonymous} codons. Although synonymous codons encode the same amino acid, they are not functionally equivalent: condon sequence matters and different organisms use them at different frequencies (a phenomenon known as \emph{codon bias}). This affects translation speed, mRNA stability, and ultimately protein expression levels~\cite{gustafsson2004codon, quax2015codon, brule2017synonymous}. When expressing a recombinant protein in a host organism, choosing codons that match the host's preferences and selecting codon sequences for optimum translation speed and stability, in a process referred to as \emph{codon optimization}, can dramatically improve expression. For a more comprehensive review of codon optimization and its biological underpinnings, we refer the reader to \cite{gustafsson2004codon} and \cite{quax2015codon}.

Codon optimization is a combinatorial optimization problem that is routinely solved in pharmaceutical and biotechnology pipelines, with direct relevance to mRNA therapeutics, including COVID-19 vaccines~\cite{polack2020mrna} and protein replacement therapies~\cite{fox2021mrna}.
We formulate the problem as a Potts model (a probabilistic graphical model with categorical variables) and compile it to an Ising model via domain-wall encoding~\cite{chancellor2019dwc}, making it executable on binary (p-bit) hardware.
We show that all three approaches, Potts sampling, Ising sampling, and a carefully tuned genetic algorithm (GA) baseline, achieve near-identical optimization quality on the SARS-CoV-2 spike protein.
The decisive differentiator is energy consumption: estimates based on validated hardware models indicate that a TSU could solve this problem using between $10^5$--$10^9$ times less energy than a conventional GPU.
All code is built on the THRML library~\cite{jelincic2025dtm} and available under an open-license at \href{https://github.com/extropic-ai/thrml}{github.com/extropic-ai/thrml}.

\section{Results}

\begin{table*}[htbp]
\centering
\caption{Optimization quality on the SARS-CoV-2 spike protein ($L = 1{,}273$) for two parameter settings (see \cref{sec:methods_codon_opt}). We report the final energy (mean $\pm$ std across 512 independent chains) and the best score found across all chains. Lower scores indicate better solutions. The number of sweeps chosen is the minimal number of sweeps that achieves a score close to the best score achieved by all methods.}
\label{tab:scores}
\vspace{2mm}
\begin{tabular}{llcc}
\toprule
Setting & Method & Mean score $\pm$ std & Best score \\
\midrule
\multirow{3}{*}{Standard (Fox)} & Potts model (10 sweeps) & $\textbf{234.6} \pm 0.2$ & 234.0 \\
 & Ising model ($2 \times 10^4$ sweeps, DWE) & $243.0 \pm 0.8$ & 240.5 \\
 & Genetic algorithm (tuned) & $234.8 \pm 0.4$ & 233.9 \\
\midrule
\multirow{3}{*}{Hard} & Potts model (10 sweeps) & $\textbf{445.2} \pm 1.0$ & 444.0 \\
 & Ising model ($4 \times 10^4$ sweeps, DWE) & $452.8 \pm 1.1$ & 449.2 \\
 & Genetic algorithm (tuned) & $446.4 \pm 1.0$ & 444.4 \\
\bottomrule
\end{tabular}
\end{table*}

\subsection{Problem setup}\label{sec:results_setup}

\begin{figure*}[p]
\centering
\includegraphics[width=\textwidth]{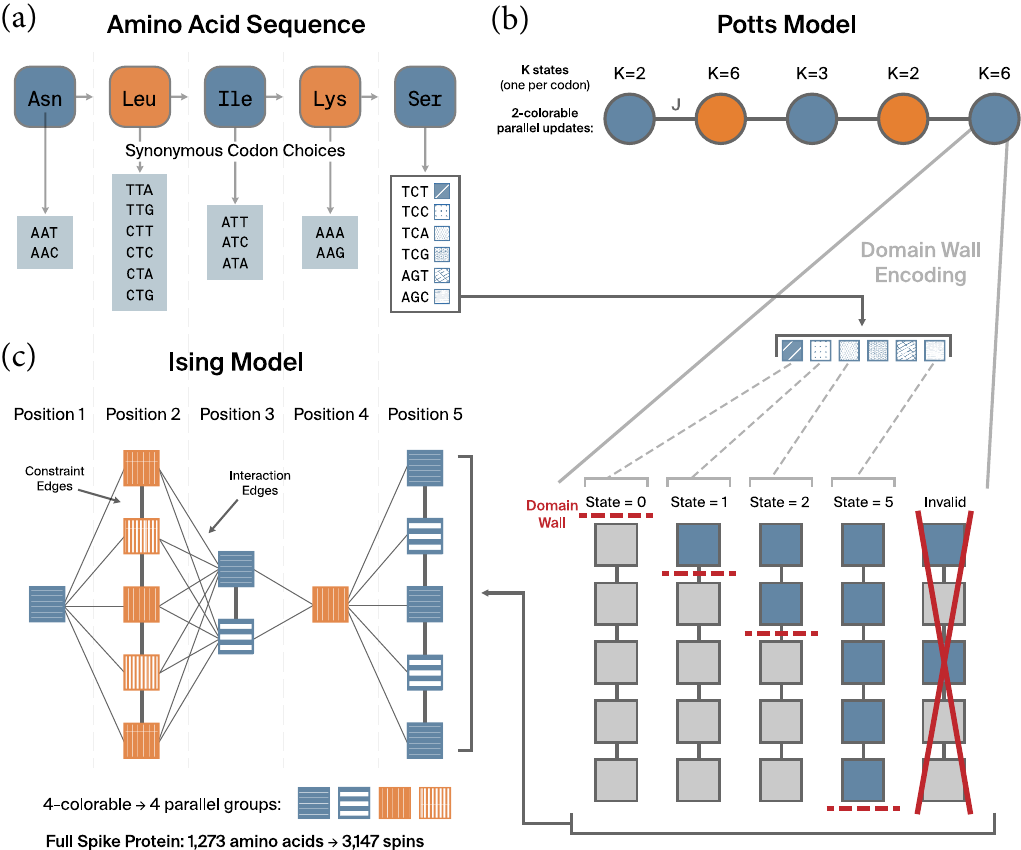}

\caption{\textbf{From codons to Ising spins.}
\textbf{(a)}~The codon optimization problem: each amino acid in a protein sequence is encoded by a triplet codon, with multiple synonymous options. The goal is to choose codons that optimize usage frequency, GC content, and repeat avoidance.
\textbf{(b)}~Potts model representation with domain-wall encoding. Each amino acid position becomes a categorical variable in a chain graph (top); the chain is 2-colorable, enabling parallel block Gibbs updates. Each $K$-state Potts variable is encoded as $K{-}1$ binary spins in a thermometer pattern (bottom), where the domain wall position encodes the categorical state.
\textbf{(c)}~Full Ising graph for a short protein segment. Intra-position constraint edges enforce validity; inter-position edges encode codon interactions. The graph is 4-colorable for parallel updates. The spike protein maps to 3,147 spins.}
\label{fig:codons-to-ising}
\end{figure*}

We benchmark all methods using the SARS-CoV-2 spike protein, a biologically relevant sequence of $L = 1{,}273$ amino acids (requiring $3{,}819$ mRNA nucleotides), using \emph{Escherichia coli} K-12 as the host organism.
The energy function to be minimized is a weighted combination of three terms (see \cref{sec:methods_codon_opt} and \cref{fig:codons-to-ising}a): a codon usage penalty ($w_f = 0.1$) that favors codons commonly used in the host, a GC content term ($w_\text{GC} = 1$) that penalizes deviation from a target GC fraction of $\rho_T = 0.5$, and a repeat penalty ($w_R = 0.1$) that discourages long runs of identical nucleotides across codon boundaries.
These weights follow the formulation of \cite{fox2021mrna} and reflect commonly used values in codon optimization practice.

We evaluate three optimization methods.
First, a \textbf{Potts model} sampler that operates directly on categorical codon variables (see \cref{sec:methods_potts} and \cref{fig:codons-to-ising}b) using simulated annealing with block Gibbs updates (10 sweeps).
Second, an \textbf{Ising model} sampler that encodes the categorical variables into binary spins via domain-wall encoding (see \cref{sec:methods_ising} and \cref{fig:codons-to-ising}b,c) and performs simulated annealing with simultaneous penalty ramping ($2 \times 10^4$ sweeps).
Third, a \textbf{genetic algorithm} (GA) baseline reimplemented from \cite{fox2021mrna} with more carefully tuned parameters (population 200, $1{,}000$ generations, mutation rate $0.003$; see \cref{sec:methods_ga} for sensitivity analysis).

\subsection{Optimization quality}\label{sec:results_quality}

\cref{tab:scores} summarizes the optimization quality achieved by each method under both the standard weights from \cite{fox2021mrna} and the harder parameter setting described in \cref{sec:methods_codon_opt}.
Under both settings, all three methods produce solutions of comparable quality.
The Potts model and GA achieve essentially identical performance, while the Ising model scores slightly higher, which is expected given the overhead of the binary encoding.

The key observation is that all methods effectively achieve the same optimization quality under both parameter settings, within ${\sim}4\%$ of each other, making the solution quality interchangeable and the energy consumption the decisive differentiator.
All energy estimates below are for the standard parameter setting from \cite{fox2021mrna}.

\subsection{Energy consumption}\label{sec:results_energy}

We estimate the energy required to solve the codon optimization problem on thermodynamic hardware and compare it to conventional GPU execution (see \cref{sec:methods_energy} for full methodology).
The TSU energy model, validated against prototype hardware measurements~\cite{jelincic2025dtm}, estimates $E_\text{cell} \approx 1.3$~fJ per spin per Gibbs step.

\paragraph{Ising chip.}
The Ising model uses $N = 3{,}147$ binary spins (p-bits) and $K = 2 \times 10^4$ Gibbs iterations.
The estimated energy is:
\begin{align}
E_\text{Ising} &= K \times N \times E_\text{cell} \nonumber \\
&= 2 \times 10^4 \times 3{,}147 \times 1.3 \times 10^{-15}~\text{J} \approx 82~\text{nJ}.
\end{align}

\paragraph{Potts chip (projected).}
Based on a physical model of categorical (p-dit) hardware extrapolated from prototype measurements, a Potts chip would consume approximately $0.084$~nJ per Gibbs sweep for $L = 1{,}273$ nodes, giving $10 \times 0.084 = 0.84$~nJ for the full optimization.
We note that Potts (p-dit) hardware does not yet exist; the Ising (p-bit) TSU is the near-term target. An example of how a Potts model could be implemented in hardware can be found in \cite{whitehead2023}. We based our energy estimates both on the measurements from our prototype and a design similar to the one proposed in \cite{whitehead2023}.

\paragraph{GPU energy.}
For GPU comparison on an NVIDIA RTX~3090 (350~W TDP), we bracket the energy between an idealized lower bound (peak FLOP throughput) and a measured upper bound (wall-clock time $\times$ TDP); see \cref{sec:methods_energy}.
The Potts model requires ${\sim}1.3 \times 10^6$ FLOPs (10 sweeps), the Ising model ${\sim}2 \times 10^9$ FLOPs ($2 \times 10^4$ sweeps), and the GA ${\sim}6 \times 10^9$ FLOPs---roughly $4{,}600\times$ more than the Potts sampler for the same solution quality.

\begin{figure}[H]
\centering
\includegraphics[width=\columnwidth]{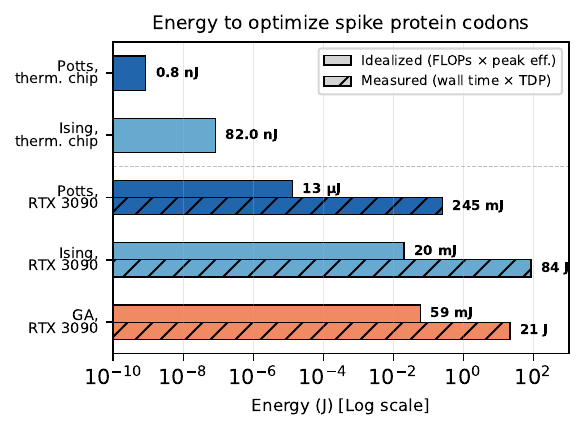}
\caption{\textbf{Energy comparison across methods and hardware.}
All methods achieve comparable optimization quality (score ${\sim}235$--$243$); the differentiator is energy consumption, which spans approximately 10 orders of magnitude.
Thermodynamic chip estimates are based on validated hardware models~\cite{jelincic2025dtm}.}
\label{fig:energy-bar}
\end{figure}

\cref{tab:energy} and \cref{fig:energy-bar} present the complete energy comparison.

\begin{table}[htbp]
\centering
\small
\caption{Energy per single-chain optimization run on the spike protein ($L = 1{,}273$), standard (Fox) weights. Ratios relative to Ising chip.}
\label{tab:energy}
\vspace{2mm}
\begin{tabular}{lrll}
\toprule
Configuration & \multicolumn{2}{c}{Energy} & Ratio v.~Ising \\
\midrule
Potts (proj.) & 0.84 & nJ & $0.01\times$ \\
Ising & 82 & nJ & $1\times$ (ref.) \\
\midrule
\multicolumn{4}{l}{\emph{GPU --- idealised (FLOPs $\times$ peak eff.)}} \\
Potts, 3090 & 13 & \textmu J & $160\times$ \\
Ising, 3090 & 20 & mJ & $2.4\!\times\!10^5$ \\
GA, 3090 & 59 & mJ & $7.2\!\times\!10^5$ \\
\midrule
\multicolumn{4}{l}{\emph{GPU --- measured (wall time $\times$ TDP)}} \\
Potts, 3090 & 245 & mJ & $3.0\!\times\!10^6$ \\
Ising, 3090 & ${\sim}84$ & J & ${\sim}10^9$ \\
GA, 3090 & 21 & J & $2.6\!\times\!10^8$ \\
\bottomrule
\end{tabular}
\end{table}

Several features of this comparison are noteworthy.
First, even at idealized peak efficiency (a generous lower bound that credits the GPU with 100\% utilization) the thermodynamic chip is between $160$ and $2.4 \times 10^5$ times more efficient than a GPU running the same algorithm (Potts and Ising respectively).
Second, realistic GPU utilization is far below peak: the Potts model's ${\sim}10^6$ FLOPs represent a tiny fraction of the RTX~3090's throughput, and execution time is dominated by kernel launch overhead.
The measured GPU energy therefore exceeds the idealized estimate by several orders of magnitude.
Third, the GA, which appears to be the most common algorithm used for codon optimization \citep{fox2021mrna}, requires ${\sim}4{,}600$ times more FLOPs than the Potts sampler for the same solution quality, compounding the hardware efficiency gap.
The combined advantage of a more efficient algorithm running on more efficient hardware yields energy savings of $10^5$--$10^9$ times relative to a GPU running the GA.

\subsection{Scaling considerations}\label{sec:results_scaling}

Solving the codon optimization problem for the spike protein requires $3{,}147$ spins in the Ising formulation, well within the capacity of near-term TSU designs (${\sim}250{,}000$ p-bits).
This leaves substantial headroom for larger proteins, longer mRNA constructs, or more complex energy functions incorporating additional biological objectives.
The chain-like graph structure of the problem, with a maximum node degree of 12, maps naturally to the sparse, locally-connected topology of a TSU.

At a random number generator (RNG) decorrelation time of ${\sim}100$~ns (measured from prototype hardware~\cite{jelincic2025dtm}), a full Gibbs step on a 4-colorable graph takes ${\sim}400$~ns.
The $2 \times 10^4$-step Ising optimization would therefore complete in approximately $8$~ms on a TSU which is more than fast enough for integration into real-time computational pipelines.

\section{Discussion}\label{sec:discussion}

This work demonstrates that a real world pharmaceutical optimization problem can be efficiently mapped to thermodynamic hardware, with energy savings of five to nine orders of magnitude compared to conventional computing (\cref{tab:energy}).
By taking a concrete pharmaceutical application from problem formulation through hardware mapping to energy estimation grounded in prototype measurements, it provides a template for evaluating thermodynamic computing on applied problems.

The Potts formulation is algorithmically superior to the Ising encoding: it uses fewer variables ($1{,}273$ categorical vs.\ $3{,}147$ binary), requires fewer sampling sweeps (10 vs.\ $2 \times 10^4$), and achieves slightly better scores.
However, building Ising (p-bit) hardware is substantially simpler than building Potts (p-dit) hardware.
A TSU capable of running the Ising formulation could be commercially available within one to two years, while categorical hardware remains further out.
Crucially, the ${\sim}100\times$ energy gap between the projected Potts chip and the Ising chip is modest compared to the $10^5$--$10^9$ times advantage that \emph{both} enjoy over GPUs.
This favors the Ising formulation as the practical near-term path.

We acknowledge several limitations:
The codon optimization objective used here is a simplified version of what production pipelines employ; more involved formulations may include terms for mRNA secondary structure, CpG dinucleotide content, and untranslated region effects~\cite{fox2021mrna}.
These can be incorporated as additional local or pairwise energy terms without fundamentally changing the graph structure or hardware requirements;
The energy estimates for the Potts chip are projections based on extrapolation from prototype measurements, while the Ising estimates rest on a validated physical model with close agreement to experimental data~\cite{jelincic2025dtm}.;
Finally, codon optimization itself is not the computational bottleneck in drug development, but it does serve as a concrete, accessible proof of concept that the pipeline from problem formulation through hardware mapping to energy estimation is viable.

More broadly, thermodynamic computing is not limited to codon optimization.
The same hardware that solves Ising models for codon sequences can accelerate conformational search, binding affinity estimation, molecular generation, and other sampling-intensive drug discovery workloads.
The key insight is that drug discovery problems naturally decompose into energy minimization over discrete or continuous variables, precisely the class of problems for which thermodynamic hardware is designed.
The rapid growth of mRNA therapeutics, from COVID-19 vaccines~\cite{polack2020mrna} to cancer immunotherapies and protein replacement therapies, ensures that codon optimization itself will remain an important step in drug development pipelines, and one where energy-efficient computation at scale becomes increasingly valuable.
As thermodynamic hardware matures, we anticipate it will unlock efficient computation across a broad range of drug discovery and similar scientific applications.

\paragraph{Code availability.}
All code for this work is open-source at \url{https://github.com/extropic-ai/codon_opt}, built on the THRML library~\cite{jelincic2025dtm}.

\section{Methods}

\subsection{Codon optimization problem}
\label{sec:methods_codon_opt}

Given a protein sequence of $L$ amino acids $a_1, \ldots, a_L$, the codon optimization problem is to select a synonymous codon $c_p \in S_p$ at each position $p$ to produce an mRNA sequence with desirable expression properties in a target host organism~\cite{gustafsson2004codon, quax2015codon, brule2017synonymous}.
The set $S_p$ contains all codons encoding amino acid $a_p$, with $|S_p| \in \{1, 2, 3, 4, 6\}$ depending on the amino acid.

The objective is to minimize the energy function:
\begin{align}
E(c_1, \ldots, c_L) &= w_f \sum_{p=1}^{L} u(c_p) + w_R \sum_{p=1}^{L-1} r(c_p, c_{p+1}) \nonumber \\
&\quad + w_\text{GC} \bigg( \frac{1}{3L} \sum_{p=1}^{L} g(c_p) - \rho_T \bigg)^{\!2}
\label{eq:energy}
\end{align}
where:
\begin{itemize}
\item The \emph{codon usage} term $u(c) = |\log(f(c) / f_\text{max}^{(a)})|$ measures how rare codon $c$ is in the host relative to the most common codon for its amino acid, using host-specific codon frequency tables~\cite{nakamura2000codon}. This term is \emph{unary}: it depends on each position independently.
\item The \emph{GC content} term penalizes deviation of the overall GC fraction from a target $\rho_T$, where $g(c) \in \{0, 1, 2, 3\}$ counts the G and C nucleotides in codon $c$. This term is \emph{global}: it couples all positions.
\item The \emph{repeat penalty} $r(c_p, c_{p+1}) = m(c_p, c_{p+1})^2 - 1$ penalizes long runs of identical nucleotides across codon boundaries, where $m$ is the length of the longest contiguous run in the 6-character concatenation. This term is \emph{pairwise}: it depends on adjacent positions only.
\end{itemize}
We follow the formulation and weights of \cite{fox2021mrna}: $w_f = 0.1$, $w_\text{GC} = 1$, and $w_R = 0.1$.
These reflect commonly used values in codon optimization practice, and all energy estimates in \cref{sec:results_energy} are based on this setting.
We additionally report optimization quality for a harder parameter setting ($w_f = 0.1$, $w_\text{GC} = 2 \times 10^4$, $w_R = 0.2$) in which the three energy terms are approximately equally balanced, making the problem substantially more difficult.
Under the standard (Fox) weights the GC content term is negligible relative to the other two, which makes the problem easier; indeed all three methods require more iterations to converge under the hard weights than under the standard ones (see \cref{tab:scores}).

\subsection{Potts model formulation}
\label{sec:methods_potts}

The codon optimization problem maps naturally to a Potts model~\cite{wu1982potts}: each amino acid position $p$ becomes a categorical random variable $X_p$ with $|S_p|$ states, one for each synonymous codon.
All variables are padded to $K_\text{max} = 6$ states, with invalid states receiving a large positive bias that effectively excludes them.

The Potts energy decomposes into unary biases $h_p(k)$ (from the codon usage term and validity constraints) and pairwise interactions $J_p(k, k')$ (from the repeat penalty between adjacent positions $p$ and $p+1$):
\begin{equation}
E(\mathbf{x}) = \sum_{p=1}^{L} h_p(x_p) + \sum_{p=1}^{L-1} J_p(x_p, x_{p+1}).
\end{equation}

Since each position interacts only with its immediate neighbors, the resulting graph is a chain, which is 2-colorable.
This enables efficient parallel block Gibbs sampling: conditioned on odd-indexed positions, all even-indexed positions are conditionally independent and can be sampled simultaneously via a softmax distribution over their $K$ states, and vice versa.
One full Gibbs sweep consists of updating the even block and then the odd block.

\subsection{Adaptive GC coefficient}\label{sec:methods_gc}

The GC content term in \cref{eq:energy} couples all positions through a global quadratic constraint.
Expanding this term directly would require all-to-all pairwise interactions, which is incompatible with the sparse connectivity of a TSU. All-to-all connectivity also increases the chromatic number of the graph to be equal to the total number of nodes, which completely prevents sampling blocks of nodes in parallel.
Instead, we approximate it with an adaptive linear term: a per-chain coefficient $\lambda$ that modifies the unary bias of each codon proportionally to its GC count:
\begin{equation}
\tilde{h}_p(k) = h_p(k) + \lambda \cdot g(c_p^{(k)}).
\end{equation}

At each annealing step, $\lambda$ is updated based on the current GC fraction error:
\begin{equation}
\lambda \leftarrow \lambda - \eta \cdot (\rho_\text{current} - \rho_T)
\end{equation}
where $\eta > 0$ is an adaptation rate.
If the current GC fraction exceeds the target, $\lambda$ decreases (penalizing high-GC codons); if below, $\lambda$ increases.
To prevent the linear approximation from dominating, $\lambda$ is clamped to never exceed the gradient of the true quadratic GC penalty, $\lambda_\text{true} = -2 w_\text{GC} (\rho_\text{current} - \rho_T) / (3L)$.

This scheme is naturally compatible with the TSU's flash-sample-read cycle: $\lambda$ is updated on the host between sampling bursts and folded into the biases before the next flash. This incurs no additional cost, since new weights would regardless need to be flashed to the TSU due to annealing.

\subsection{Ising embedding via domain-wall encoding}
\label{sec:methods_ising}

Binary spin hardware (with p-bits) is simpler to build and closer to commercial availability than categorical hardware (with p-dits).
To make the codon optimization problem executable on a binary TSU, we compile the Potts model into an Ising model using domain-wall encoding (DWE)~\cite{chancellor2019dwc, berwald2022dwc, chen2021dwc}.

A $K$-state Potts variable is represented by $K{-}1$ binary spins $s_1, \ldots, s_{K-1} \in \{-1, +1\}$ arranged in a chain.
Valid states follow a thermometer pattern: state $k$ corresponds to $s_1 = \cdots = s_k = +1$ and $s_{k+1} = \cdots = s_{K-1} = -1$.
The ``domain wall''---the boundary between the $+1$ and $-1$ regions---encodes the Potts state.
Invalid configurations (with a $-1$ followed by a $+1$) are penalized by ferromagnetic nearest-neighbor couplings of strength $P/4$ within each position's spin chain, adding $K{-}2$ constraint edges per position.

DWE offers several advantages over the more common one-hot encoding~\cite{lucas2014ising}.
First, the constraint graph is a chain ($K{-}2$ edges) rather than a clique ($K(K{-}1)/2$ edges), yielding a much sparser graph.
Second, the chain constraint is 2-colorable, which at most doubles the chromatic number of the overall graph---unlike one-hot, which requires $K$ colors for its penalty clique, which can therefore increase the chromatic number of the graph $K$-fold. Since the chromatic number of the graph determines how many sampling steps are required per Gibbs iteration, the one-hot encoding is typically $K/2$-times slower than DWE per Gibbs iteration, on top of possibly also requiring a far greater number of Gibbs iterations due to its slow mixing.
Third, mixing at the domain wall boundary is independent of the penalty strength $P$: the ferromagnetic contributions from the two neighbors cancel, so the Gibbs conditional depends only on the bias landscape.
This eliminates the exponential mixing slowdown~\cite{levin2009markov} that plagues one-hot encoding at large $P$~\cite{berwald2022dwc}.
Fourth, DWE has been proven optimal in the number of binary variables for quadratic interactions~\cite{berwald2022dwc}.

Potts biases are compiled to Ising fields via first differences ($b_i = (h_i - h_{i-1})/2$ plus boundary corrections), and Potts pairwise interactions become Ising couplings via mixed second differences ($J_{(p,i),(q,j)} = (V_{ij} - V_{i-1,j} - V_{i,j-1} + V_{i-1,j-1})/4$); see Supplementary Note~1 for the full derivation.

\paragraph{GC-sorted codon ordering.}
In DWE, a single spin flip moves the Potts variable to an adjacent state ($k \to k \pm 1$).
We order codons at each position by ascending GC count (with alphabetical tiebreaking), so that a single spin flip changes GC content by at most 1.
This creates a smooth landscape for domain wall dynamics and ensures compatibility with the adaptive GC scheme.

The resulting Ising model for the spike protein contains $N = 3{,}147$ spins (an average of $2.47$ per amino acid position), with a maximum node degree of 12 and only local connectivity.
The graph is 4-colorable using a decomposition based on position parity and spin-index parity.

\subsection{Simulated annealing and P-ramping}\label{sec:methods_annealing}

Both the Potts and Ising models use simulated annealing with a log-spaced inverse temperature schedule:
\begin{equation}
\beta_t = \beta_\text{min} \cdot \left( \frac{\beta_\text{max}}{\beta_\text{min}} \right)^{t/T}
\end{equation}
for $t \in \{0, \ldots, T\}$.
At each temperature step, the host updates $\beta$, recomputes the adaptive GC coefficient, flashes new weights to the TSU, and runs a fixed number of Gibbs sweeps.

For the Ising model, we simultaneously ramp the constraint penalty $P$ on an independent log-spaced schedule from $P_\text{min}$ to $P_\text{max}$.
Low $P$ early in annealing permits invalid-state (defect) transitions that provide shortcuts around energy barriers in the bias landscape: when a valid intermediate state has energy exceeding $P$, the defect route is energetically cheaper.
High $P$ at the end enforces strict constraint satisfaction.
Since weights are recomputed at each annealing step anyway (for $\beta$ and the adaptive GC coefficient), changing $P$ adds no extra cost.

\subsection{Genetic algorithm baseline}\label{sec:methods_ga}

Genetic algorithms are widely used for codon optimization~\cite{fox2021mrna, sandhu2008gasco}.
As a baseline, we reimplemented the GA from \cite{fox2021mrna} in JAX, with full JIT compilation of the generation loop and multi-chain parallelism.
After systematic parameter tuning, the best configuration uses a population of 200, $1{,}000$ generations, 10 elite and 2 lucky survivors per generation, and a mutation rate of $0.003$.

The GA's performance is sensitive to parameter choice.
The default mutation rate of $0.05$ from Fox et al.\ yields substantially worse scores (${\sim}550$--$600$), suggesting that the crossover operator provides sufficient exploration and high mutation rates are counterproductive, disrupting good partial solutions faster than selection can preserve them.

\subsection{Energy estimation methodology}\label{sec:methods_energy}

\paragraph{TSU energy model.}
We use the energy model from \cite{jelincic2025dtm} (Appendix E), which provides a physical model of an all-transistor Boltzmann machine Gibbs sampler.
The energy per spin per Gibbs step is $E_\text{cell} \approx 1.3$~fJ, which includes contributions from the RNG (${\sim}350$~aJ), biasing circuitry, clocking, and inter-cell communication.
This model captures all central functional units of the hardware and has been validated against measurements from a prototype chip, with agreement within an order of magnitude.

For the Ising chip, the total energy is $E = K \times N \times E_\text{cell}$, where $K$ is the number of Gibbs iterations and $N$ is the number of spins.
For the Potts chip, we use a physical model of p-dit hardware extrapolated from the same prototype measurements, yielding ${\sim}0.084$~nJ per Gibbs sweep for $L = 1{,}273$ Potts nodes.

Our Ising model has a maximum degree of 12 with only local connectivity, closely matching the setup analyzed in \cite{jelincic2025dtm}.
The codon problem ($3{,}147$ spins) fits comfortably within the capacity of near-term TSU designs (${\sim}250{,}000$ p-bits).

\paragraph{GPU energy model.}
We estimate GPU energy on an NVIDIA RTX~3090 (350~W TDP, 35.6~TFLOPS FP32 peak) using two methods.
The \emph{idealized} estimate computes total FLOPs divided by peak FLOP/s, multiplied by TDP.
This represents a lower bound, as no real workload sustains 100\% utilization.
The \emph{measured} estimate multiplies the wall-clock time of our JIT-compiled JAX implementation by TDP, giving an upper bound since the GPU may not draw full TDP power during execution.
FLOP counts were estimated by analyzing the per-step computational requirements of each algorithm (see Supplementary Note~2 for details).

\begin{strip}
\vspace{0.8em}
\noindent\rule{\textwidth}{1pt}
\begin{center}\large\bfseries References\end{center}
\vspace{-0.3em}
\end{strip}
\renewcommand{\section}[2]{}
{\small \bibliography{refs}}

@techreport{iea2024,
  author       = {{International Energy Agency}},
  title        = {Electricity 2024: Analysis and Forecast to 2026},
  institution  = {IEA},
  address      = {Paris},
  year         = {2024},
  url          = {https://www.iea.org/reports/electricity-2024},
  note         = {Licence: CC BY 4.0}
}

@article{hooker2021hardware,
  author    = {Hooker, Sara},
  title     = {The Hardware Lottery},
  journal   = {Communications of the ACM},
  volume    = {64},
  number    = {12},
  pages     = {58--65},
  year      = {2021},
  month     = nov,
  doi       = {10.1145/3467017},
  issn      = {1557-7317},
  publisher = {Association for Computing Machinery (ACM)}
}

@misc{jelincic2025dtm,
  author    = {Jelin\v{c}i\v{c}, Andra\v{z} and Lockwood, Owen and Garlapati, Akhil and Schillinger, Peter and Chuang, Isaac and Verdon, Guillaume and McCourt, Trevor},
  title     = {An efficient probabilistic hardware architecture for diffusion-like models},
  year      = {2025},
  eprint    = {2510.23972},
  archiveprefix = {arXiv},
  primaryclass  = {cs.LG},
  doi       = {10.48550/arXiv.2510.23972},
  url       = {https://arxiv.org/abs/2510.23972}
}

@misc{hylton2019thermo,
  author    = {Conte, Tom and DeBenedictis, Erik and Ganesh, Natesh and Hylton, Todd and Strachan, John Paul and Williams, R. Stanley and Alemi, Alexander and Altenberg, Lee and Crooks, Gavin and Crutchfield, James and del Rio, Lidia and Deutsch, Josh and DeWeese, Michael and Douglas, Khari and Esposito, Massimiliano and Frank, Michael and Fry, Robert and Harsha, Peter and Hill, Mark and Kello, Christopher and Krichmar, Jeff and Kumar, Suhas and Liu, Shih-Chii and Lloyd, Seth and Marsili, Matteo and Nemenman, Ilya and Nugent, Alex and Packard, Norman and Randall, Dana and Sadowski, Peter and Santhanam, Narayana and Shaw, Robert and Stieg, Adam and Stopnitzky, Elan and Teuscher, Christof and Watkins, Chris and Wolpert, David and Yang, Joshua and Yufik, Yan},
  title     = {Thermodynamic Computing},
  year      = {2019},
  eprint    = {1911.01968},
  archiveprefix = {arXiv},
  primaryclass  = {cs.CY},
  doi       = {10.48550/arXiv.1911.01968},
  url       = {https://arxiv.org/abs/1911.01968}
}

@inproceedings{hylton2020thermo,
  author    = {Hylton, Todd},
  title     = {Thermodynamic Computing: An Intellectual and Technological Frontier},
  booktitle = {IS4SI 2019 Summit},
  series    = {Proceedings},
  volume    = {47},
  pages     = {23},
  year      = {2020},
  month     = jun,
  doi       = {10.3390/proceedings2020047023},
  publisher = {MDPI}
}

@article{fox2021mrna,
  author    = {Fox, Dillion M. and Branson, Kim M. and Walker, Ross C.},
  editor    = {Chancellor, Nicholas},
  title     = {m{RNA} codon optimization with quantum computers},
  journal   = {PLOS ONE},
  volume    = {16},
  number    = {10},
  pages     = {e0259101},
  year      = {2021},
  month     = oct,
  doi       = {10.1371/journal.pone.0259101},
  issn      = {1932-6203},
  publisher = {Public Library of Science (PLoS)}
}

@article{gustafsson2004codon,
  author    = {Gustafsson, Claes and Govindarajan, Sridhar and Minshull, Jeremy},
  title     = {Codon bias and heterologous protein expression},
  journal   = {Trends in Biotechnology},
  volume    = {22},
  number    = {7},
  pages     = {346--353},
  year      = {2004},
  month     = jul,
  doi       = {10.1016/j.tibtech.2004.04.006},
  issn      = {0167-7799},
  publisher = {Elsevier BV}
}

@article{quax2015codon,
  author    = {Quax, Tessa E. F. and Claassens, Nico J. and S{\"o}ll, Dieter and van der Oost, John},
  title     = {Codon Bias as a Means to Fine-Tune Gene Expression},
  journal   = {Molecular Cell},
  volume    = {59},
  number    = {2},
  pages     = {149--161},
  year      = {2015},
  month     = jul,
  doi       = {10.1016/j.molcel.2015.05.035},
  issn      = {1097-2765},
  publisher = {Elsevier BV}
}

@article{brule2017synonymous,
  author    = {Brul{\'e}, Christina E. and Grayhack, Elizabeth J.},
  title     = {Synonymous Codons: Choose Wisely for Expression},
  journal   = {Trends in Genetics},
  volume    = {33},
  number    = {4},
  pages     = {283--297},
  year      = {2017},
  month     = apr,
  doi       = {10.1016/j.tig.2017.02.001},
  issn      = {0168-9525},
  publisher = {Elsevier BV}
}

@article{nakamura2000codon,
  author    = {Nakamura, Y. and Gojobori, T. and Ikemura, T.},
  title     = {Codon usage tabulated from international {DNA} sequence databases: status for the year 2000},
  journal   = {Nucleic Acids Research},
  volume    = {28},
  number    = {1},
  pages     = {292},
  year      = {2000},
  month     = jan,
  doi       = {10.1093/nar/28.1.292},
  issn      = {1362-4962},
  publisher = {Oxford University Press (OUP)}
}

@article{sandhu2008gasco,
  author    = {Sandhu, Kuljeet Singh and Pandey, Sunil and Maiti, Souvik and Pillai, Beena},
  title     = {{GASCO}: Genetic Algorithm Simulation for Codon Optimization},
  journal   = {In Silico Biology},
  volume    = {8},
  number    = {2},
  pages     = {187--192},
  year      = {2008},
  doi       = {10.3233/ISB-00354},
  issn      = {1434-3207},
  publisher = {SAGE Publications}
}

@article{polack2020mrna,
  author    = {Polack, Fernando P. and Thomas, Stephen J. and Kitchin, Nicholas and Absalon, Judith and Gurtman, Alejandra and Lockhart, Stephen and Perez, John L. and P{\'e}rez Marc, Gonzalo and Moreira, Edson D. and Zerbini, Cristiano and Bailey, Ruth and Swanson, Kena A. and Roychoudhury, Satrajit and Koury, Kenneth and Li, Ping and Kalina, Warren V. and Cooper, David and Frenck, Robert W. and Hammitt, Laura L. and T{\"u}reci, {\"O}zlem and Nell, Haylene and Schaefer, Axel and {\"U}nal, Serhat and Tresnan, Dina B. and Mather, Susan and Dormitzer, Philip R. and \c{S}ahin, U\u{g}ur and Jansen, Kathrin U. and Gruber, William C.},
  title     = {Safety and Efficacy of the {BNT162b2} m{RNA} {C}ovid-19 Vaccine},
  journal   = {New England Journal of Medicine},
  volume    = {383},
  number    = {27},
  pages     = {2603--2615},
  year      = {2020},
  month     = dec,
  doi       = {10.1056/NEJMoa2034577},
  issn      = {1533-4406},
  publisher = {Massachusetts Medical Society}
}

@article{chancellor2019dwc,
  author    = {Chancellor, Nicholas},
  title     = {Domain wall encoding of discrete variables for quantum annealing and {QAOA}},
  journal   = {Quantum Science and Technology},
  volume    = {4},
  number    = {4},
  pages     = {045004},
  year      = {2019},
  month     = aug,
  doi       = {10.1088/2058-9565/ab33c2},
  issn      = {2058-9565},
  publisher = {IOP Publishing}
}

@article{berwald2022dwc,
  author    = {Berwald, Jesse and Chancellor, Nicholas and Dridi, Raouf},
  title     = {Understanding domain-wall encoding theoretically and experimentally},
  journal   = {Philosophical Transactions of the Royal Society A: Mathematical, Physical and Engineering Sciences},
  volume    = {381},
  number    = {2241},
  year      = {2022},
  month     = dec,
  doi       = {10.1098/rsta.2021.0410},
  issn      = {1471-2962},
  publisher = {The Royal Society}
}

@article{chen2021dwc,
  author    = {Chen, Jie and Stollenwerk, Tobias and Chancellor, Nicholas},
  title     = {Performance of Domain-Wall Encoding for Quantum Annealing},
  journal   = {IEEE Transactions on Quantum Engineering},
  volume    = {2},
  pages     = {1--14},
  year      = {2021},
  doi       = {10.1109/TQE.2021.3094280},
  issn      = {2689-1808},
  publisher = {Institute of Electrical and Electronics Engineers (IEEE)}
}

@article{lucas2014ising,
  author    = {Lucas, Andrew},
  title     = {Ising formulations of many {NP} problems},
  journal   = {Frontiers in Physics},
  volume    = {2},
  pages     = {5},
  year      = {2014},
  doi       = {10.3389/fphy.2014.00005},
  issn      = {2296-424X},
  publisher = {Frontiers Media SA}
}

@article{wu1982potts,
  author    = {Wu, F. Y.},
  title     = {The {P}otts model},
  journal   = {Reviews of Modern Physics},
  volume    = {54},
  number    = {1},
  pages     = {235--268},
  year      = {1982},
  month     = jan,
  doi       = {10.1103/RevModPhys.54.235},
  issn      = {0034-6861},
  publisher = {American Physical Society (APS)}
}

@book{levin2009markov,
  author    = {Levin, David A. and Peres, Yuval and Wilmer, Elizabeth L.},
  title     = {Markov Chains and Mixing Times},
  publisher = {American Mathematical Society},
  address   = {Providence, RI},
  year      = {2009},
  isbn      = {978-0-8218-4739-8}
}

@article{kirkpatrick1983sa,
  author    = {Kirkpatrick, S. and Gelatt, C. D. and Vecchi, M. P.},
  title     = {Optimization by Simulated Annealing},
  journal   = {Science},
  volume    = {220},
  number    = {4598},
  pages     = {671--680},
  year      = {1983},
  month     = may,
  doi       = {10.1126/science.220.4598.671},
  issn      = {1095-9203},
  publisher = {American Association for the Advancement of Science (AAAS)}
}

@article{whitehead2023,
  author    = {Whitehead, William and Nelson, Zachary and Camsari, Kerem Y. and Theogarajan, Luke},
  title     = {{CMOS}-compatible {I}sing and {P}otts annealing using single-photon avalanche diodes},
  journal   = {Nature Electronics},
  volume    = {6},
  number    = {12},
  pages     = {1009--1019},
  year      = {2023},
  month     = nov,
  doi       = {10.1038/s41928-023-01065-0},
  issn      = {2520-1131},
  publisher = {Springer Science and Business Media LLC}
}

\end{document}


\maketitle

\section*{Supplementary Note 1: Domain-wall encoding derivation}

This note provides the full mathematical derivation of how Potts model weights are compiled to Ising model weights via domain-wall encoding (DWE)~\cite{chancellor2019dwc, berwald2022dwc}, as referenced in Section~4.4 of the main text.

\subsection*{Setup and notation}

Consider a single Potts variable $c \in \{0, 1, \ldots, K{-}1\}$ with energy $E(c) = h_c$ for each state.
We seek an Ising model on $\{-1,+1\}$ spins that reproduces these energies exactly (up to a constant) on valid configurations and penalizes invalid ones.

\subsection*{Thermometer encoding and constraint}

Introduce $K{-}1$ Ising spins $s_1, \ldots, s_{K-1} \in \{-1,+1\}$.
Potts state $k$ is encoded as the ``thermometer'' pattern~\cite{chancellor2019dwc} where the first $k$ spins are $+1$ and the rest are $-1$:
\begin{equation}\label{eq:thermo}
  c = k \quad\Longleftrightarrow\quad
  s_i = \begin{cases} +1 & i \leq k \\ -1 & i > k \end{cases}
\end{equation}
The ``domain wall''---the boundary between the $+1$ and $-1$ regions---sits at position $k$.

Invalid configurations (those with a $-1$ followed by a $+1$, i.e.\ a reverse domain wall) are penalized.
In $\{0,1\}$ variables $q_i = (1+s_i)/2$, the penalty for a single reverse wall between positions $i$ and $i{+}1$ is $P \cdot q_{i+1}(1 - q_i)$.
Converting to $\{-1,+1\}$ spins:
\begin{equation}\label{eq:constraint}
  H_{\mathrm{constr}} = \frac{P}{4} \sum_{i=1}^{K-2} \bigl(1 - s_i + s_{i+1} - s_i s_{i+1}\bigr)
\end{equation}
This equals zero on all valid (thermometer) configurations and adds $P$ for each reverse domain wall.
The constraint graph is a \emph{chain} ($K{-}2$ edges), which is 2-colorable---in contrast to the $K(K{-}1)/2$-edge clique required by one-hot encoding~\cite{lucas2014ising}.

\subsection*{Single-variable Ising Hamiltonian}

The indicator for Potts state $k$ can be written as $\delta_{c=k} = q_k - q_{k+1}$ (defining $q_0 \equiv 1$, $q_K \equiv 0$).
Therefore $\sum_k h_k\,\delta_{c=k} = h_0 + \sum_{i=1}^{K-1}(h_i - h_{i-1})\,q_i$.
Converting to $\{-1,+1\}$ spins:
\begin{equation}\label{eq:bias}
  H_{\mathrm{bias}} = \sum_{i=1}^{K-1} \frac{h_i - h_{i-1}}{2}\, s_i + \mathrm{const}
\end{equation}

Combining \cref{eq:constraint} and \cref{eq:bias} and collecting terms:
\begin{equation}\label{eq:single-full}
\boxed{
  H = \sum_{i=1}^{K-1} b_i\, s_i \;-\; \frac{P}{4}\sum_{i=1}^{K-2} s_i\, s_{i+1} \;+\; \mathrm{const}
}
\end{equation}
where the local fields are:
\begin{equation}\label{eq:fields}
  b_i = \frac{h_i - h_{i-1}}{2} +
  \begin{cases}
    -P/4 & i = 1 \\
    0 & 2 \leq i \leq K{-}2 \\
    +P/4 & i = K{-}1
  \end{cases}
\end{equation}
The couplings $J_{i,i+1} = -P/4$ are \textbf{ferromagnetic} (favoring aligned spins), forming a nearest-neighbor chain.

\subsection*{Worked example: $K=4$}

Consider Potts biases $h_0, h_1, h_2, h_3$.
Three Ising spins $s_1, s_2, s_3$ yield the Hamiltonian:
\begin{equation}
  H = b_1\, s_1 + b_2\, s_2 + b_3\, s_3 - \tfrac{P}{4}\, s_1 s_2 - \tfrac{P}{4}\, s_2 s_3 + \mathrm{const}
\end{equation}
with
\begin{equation}
  b_1 = \frac{h_1 - h_0}{2} - \frac{P}{4}, \qquad
  b_2 = \frac{h_2 - h_1}{2}, \qquad
  b_3 = \frac{h_3 - h_2}{2} + \frac{P}{4}
\end{equation}

\subsection*{Multi-variable case: pairwise interactions}

Consider $L$ Potts variables $c_1, \ldots, c_L$ with unary biases $h^{(p)}_k$ and pairwise interactions $V^{(pq)}_{kl}$~\cite{chancellor2019dwc, berwald2022dwc}:
\begin{equation}\label{eq:potts-energy}
  E = \sum_p \sum_k h^{(p)}_k \,\delta_{c_p=k}
    \;+\; \sum_{(p,q)} \sum_{k,l} V^{(pq)}_{kl}\,\delta_{c_p=k}\,\delta_{c_q=l}
\end{equation}
Each position $p$ gets its own domain-wall chain of $K_p{-}1$ spins $s_{p,1},\ldots,s_{p,K_p-1}$.
The unary terms convert exactly as before (\cref{eq:bias,eq:fields}).
Below we derive how pairwise Potts interactions compile to Ising couplings.

\paragraph{Step 1: Abel summation in $\{0,1\}$ variables.}

Using the indicator identity $\delta_{c_p=k} = q_{p,k} - q_{p,k+1}$ (with $q_{p,0}\equiv 1$, $q_{p,K_p}\equiv 0$) and the standard Abel summation formula
$\sum_{k=0}^{K-1} f_k\,(q_k - q_{k+1}) = f_0 + \sum_{i=1}^{K-1}(f_i - f_{i-1})\,q_i$,
we expand the pairwise energy between positions $p$ and $q$.

First, sum over $l$ for fixed $k$:
\begin{equation}
  \sum_l V_{kl}\,(q_{q,l} - q_{q,l+1})
  = V_{k,0} + \sum_{j=1}^{K_q-1}(V_{k,j} - V_{k,j-1})\,q_{q,j}
\end{equation}
Then sum over $k$, applying Abel summation to each coefficient separately:
\begin{equation}\label{eq:abel-result}
  E_{pq}
  = V_{0,0}
    + \sum_{i=1}^{K_p-1} \alpha_i\, q_{p,i}
    + \sum_{j=1}^{K_q-1} \beta_j\, q_{q,j}
    + \sum_{i=1}^{K_p-1}\sum_{j=1}^{K_q-1} \widetilde{V}_{ij}\, q_{p,i}\, q_{q,j}
\end{equation}
where
\begin{align}
  \alpha_i &= V_{i,0} - V_{i-1,0}
    & &\text{(first difference along column $l{=}0$)} \label{eq:alpha}\\
  \beta_j &= V_{0,j} - V_{0,j-1}
    & &\text{(first difference along row $k{=}0$)} \label{eq:beta}\\[4pt]
  \widetilde{V}_{ij} &= V_{ij} - V_{i-1,j} - V_{i,j-1} + V_{i-1,j-1}
    & &\text{(mixed second difference)} \label{eq:mixed-diff}
\end{align}

\paragraph{Step 2: convert to $\{-1,+1\}$ spins.}

Substitute $q_{p,i} = (1+s_{p,i})/2$ into \cref{eq:abel-result}.
The linear and quadratic terms expand as:
\begin{align}
  \alpha_i\, q_{p,i} &= \tfrac{\alpha_i}{2} + \tfrac{\alpha_i}{2}\, s_{p,i} \\[2pt]
  \widetilde{V}_{ij}\, q_{p,i}\, q_{q,j}
    &= \tfrac{\widetilde{V}_{ij}}{4}
     + \tfrac{\widetilde{V}_{ij}}{4}\, s_{p,i}
     + \tfrac{\widetilde{V}_{ij}}{4}\, s_{q,j}
     + \tfrac{\widetilde{V}_{ij}}{4}\, s_{p,i}\, s_{q,j}
\end{align}
Collecting all terms by type yields:

\paragraph{Inter-position couplings.}
\begin{equation}\label{eq:Jinter}
\boxed{\;
  J^{\mathrm{inter}}_{(p,i),(q,j)} = \frac{\widetilde{V}_{ij}}{4}
  = \frac{V_{ij} - V_{i-1,j} - V_{i,j-1} + V_{i-1,j-1}}{4}
\;}
\end{equation}

\paragraph{Bias contributions from pairwise interactions.}
Each pairwise interaction $(p,q)$ contributes an additional bias to spins at both positions.
For spin $s_{p,i}$:
\begin{equation}\label{eq:bias-pairwise}
  b_{p,i}^{(pq)}
  = \frac{\alpha_i}{2} + \frac{1}{4}\sum_{j=1}^{K_q-1} \widetilde{V}_{ij}
  = \frac{(V_{i,0} - V_{i-1,0}) + (V_{i,K_q-1} - V_{i-1,K_q-1})}{4}
\end{equation}
where the simplification uses the fact that $\sum_j \widetilde{V}_{ij}$ telescopes.
Symmetrically, for spin $s_{q,j}$:
\begin{equation}\label{eq:bias-pairwise-q}
  b_{q,j}^{(pq)}
  = \frac{(V_{0,j} - V_{0,j-1}) + (V_{K_p-1,j} - V_{K_p-1,j-1})}{4}
\end{equation}

\paragraph{Complete multi-position Ising Hamiltonian.}
Combining unary terms (\cref{eq:fields}), constraint chains (\cref{eq:constraint}), and pairwise terms (\cref{eq:Jinter,eq:bias-pairwise,eq:bias-pairwise-q}):
\begin{equation}\label{eq:full-multi}
\boxed{
  H = \sum_p \sum_{i=1}^{K_p-1} b_{p,i}\, s_{p,i}
    \;-\; \frac{P}{4}\sum_p \sum_{i=1}^{K_p-2} s_{p,i}\, s_{p,i+1}
    \;+\; \sum_{(p,q)} \sum_{i,j} \frac{\widetilde{V}^{(pq)}_{ij}}{4}\, s_{p,i}\, s_{q,j}
    \;+\; \mathrm{const}
}
\end{equation}
The total bias on spin $s_{p,i}$ is:
\begin{equation}\label{eq:total-bias}
  b_{p,i} = \underbrace{\frac{h^{(p)}_i - h^{(p)}_{i-1}}{2}}_{\text{Potts unary}}
  + \underbrace{\text{boundary}_{P}(i)}_{\text{constraint}}
  + \underbrace{\sum_{q:\,(p,q)\in\mathcal{E}} b_{p,i}^{(pq)}}_{\text{pairwise (\cref{eq:bias-pairwise})}}
\end{equation}
where $\text{boundary}_P(i) = -P/4$ if $i=1$, $+P/4$ if $i=K_p{-}1$, and $0$ otherwise.

The Ising model has two types of edges:
\begin{itemize}
  \item \textbf{Constraint chains} (coupling $-P/4$, ferromagnetic): $s_{p,i}\leftrightarrow s_{p,i+1}$ within each position.
  \item \textbf{Inter-position couplings} (coupling $\widetilde{V}_{ij}/4$): $s_{p,i}\leftrightarrow s_{q,j}$ for each pair of interacting Potts positions, with $(K_p{-}1)(K_q{-}1)$ such couplings per Potts edge.
\end{itemize}

\subsection*{P-independent mixing at the domain wall}

A Gibbs update at the domain wall boundary---where spin $s_i$ has neighbor $s_{i-1} = +1$ (inside the wall) and $s_{i+1} = -1$ (outside)---sees an effective field:
\begin{equation}
  h_{\mathrm{eff}} = b_i + J \cdot s_{i-1} + J \cdot s_{i+1} = b_i + J \cdot (+1) + J \cdot (-1) = b_i
\end{equation}
The two ferromagnetic coupling contributions cancel exactly.
The domain wall therefore performs a random walk governed only by the bias landscape $\{b_i\}$, completely independent of the penalty strength $P$.

This is the key advantage of DWE over one-hot encoding.
In one-hot encoding, the penalty clique creates genuine energy barriers that scale with $P$, leading to exponentially slow mixing at large $P$~\cite{berwald2022dwc, levin2009markov}.
In DWE, the constraint can be made arbitrarily strong without affecting the sampling dynamics at the domain wall.

\subsection*{4-color block decomposition}

The Ising graph for the codon problem is 4-colorable using the coloring $(p_{\mathrm{parity}},\, j_{\mathrm{parity}})$, where $p_{\mathrm{parity}} = p \bmod 2$ is the position parity and $j_{\mathrm{parity}} = j \bmod 2$ is the spin-index parity within each position's chain.
The key observation is that constraint edges connect spins within the same position (same $p_{\mathrm{parity}}$, different $j_{\mathrm{parity}}$), while inter-position edges connect spins at adjacent positions (different $p_{\mathrm{parity}}$).
Therefore, no two spins sharing the same $(p_{\mathrm{parity}},\, j_{\mathrm{parity}})$ pair are connected by an edge, and each of the four color classes can be updated simultaneously in a parallel Gibbs step.

\subsection*{Comparison of encoding schemes}

\Cref{tab:encodings} summarizes the key properties of three encoding schemes for mapping a $K$-state Potts variable to binary spins.

\begin{table}[H]
\centering
\caption{Comparison of encoding schemes for a $K$-state Potts variable. Values in parentheses are for $K=6$ (the maximum in the codon problem).}
\label{tab:encodings}
\vspace{2mm}
\begin{tabular}{lcccc}
\toprule
Property & Native Potts & One-hot & Domain wall & Binary \\
\midrule
Variables & 1 categorical & $K$ (6) & $K{-}1$ (5) & $\lceil\log_2 K\rceil$ (3) \\
Constraint edges & 0 & $\frac{K(K{-}1)}{2}$ (15) & $K{-}2$ (4) & complex \\
Penalty strength & N/A & large & ${\sim}2$--$5\times$ smaller & complex \\
Chromatic impact & none & $\geq K$-fold & none (2-colorable) & unpredictable \\
Gibbs mixing & optimal & poor at large $P$ & $P$-independent & very poor \\
\bottomrule
\end{tabular}
\end{table}

Domain-wall encoding offers several advantages over one-hot encoding~\cite{chancellor2019dwc, berwald2022dwc, chen2021dwc}.
The constraint graph is a chain rather than a clique, yielding a much sparser graph and lower penalty requirements.
The chain constraint is 2-colorable, so it at most doubles the chromatic number of the overall graph---unlike one-hot, which can increase it $K$-fold.
DWE has been proven optimal in the number of binary variables for quadratic interactions~\cite{berwald2022dwc}.
Finally, Chen et al.~\cite{chen2021dwc} demonstrated experimentally that DWE outperforms one-hot encoding, with a D-Wave 2000Q using DWE outperforming the next-generation D-Wave Advantage with one-hot encoding.
Binary encoding uses the fewest variables ($\lceil\log_2 K\rceil$) but introduces complex, non-local constraint structures that are difficult to enforce and lead to poor mixing.

\section*{Supplementary Note 2: Energy estimation details}

This note provides the detailed methodology for the energy estimates presented in Section~4.7 of the main text.

\subsection*{TSU energy model (Ising chip)}

We use the energy model from Jelincic et al.~\cite{jelincic2025dtm} (Appendix~E), which provides a physical model of an all-transistor Boltzmann machine Gibbs sampler.
The total energy for a sampling run is:
\begin{equation}\label{eq:ising-energy}
  E = n_{\mathrm{iter}} \times N \times E_{\mathrm{cell}}
\end{equation}
where $n_{\mathrm{iter}}$ is the number of Gibbs iterations, $N$ is the number of spins, and $E_{\mathrm{cell}} \approx 1.3$~fJ is the energy per spin per Gibbs step.

The per-cell energy $E_{\mathrm{cell}}$ breaks down into four components:
\begin{itemize}
  \item $E_{\mathrm{rng}} \approx 350$~aJ: energy consumed by the all-transistor random number generator, based on direct measurements from a prototype chip. The RNG exploits shot-noise dynamics of subthreshold transistors and has a decorrelation time of ${\sim}100$~ns.
  \item $E_{\mathrm{bias}}$: energy consumed by the biasing circuitry (a linear resistor network that computes a biasing voltage from neighbor states).
  \item $E_{\mathrm{clock}}$: energy consumed by clocking circuitry that orchestrates the sequential color-block updates.
  \item $E_{\mathrm{comm}}$: energy consumed by inter-cell communication (charging wires to neighboring cells).
\end{itemize}

This model captures all central functional units of a hardware Boltzmann machine sampler.
When comparing the results of this type of calculation to a detailed analysis of a complete device design (Cadence simulations), agreement is found within an order of magnitude~\cite{jelincic2025dtm}.

For the codon optimization problem, the Ising model has $N = 3{,}147$ spins with a maximum degree of 12 and only local connectivity, closely matching the setup analyzed in~\cite{jelincic2025dtm}.
With $n_{\mathrm{iter}} = 2 \times 10^4$ Gibbs iterations:
\begin{equation}
  E_{\mathrm{Ising}} = 2 \times 10^4 \times 3{,}147 \times 1.3 \times 10^{-15}~\text{J} \approx 82~\text{nJ}
\end{equation}

\subsection*{TSU energy model (Potts chip, projected)}

The Potts chip energy estimate is based on a physical model of a p-dit (categorical) sampling circuit, extrapolated from prototype measurements of p-bit circuits.
The key energy-consuming components are:

\begin{itemize}
  \item \textbf{P-bit sampling}: each p-cell draws ${\sim}800$~aJ per random sample from its internal stochastic circuit.
  \item \textbf{MACDAC} (multiply-accumulate DAC): computes the conditional bias for each Potts state from neighbor states. For a node with $q$ states and $d$ neighbors, the MACDAC has $d \cdot \lceil\log_2 q\rceil + 2$ inputs, with an output capacitance of ${\sim}500$~aF per input.
  \item \textbf{Communication}: wires carrying state information to neighboring cells, with capacitance ${\sim}350$~aF/$\mu$m.
  \item \textbf{Clocking}: global clock distribution for coordinating block updates.
\end{itemize}

Each p-cell requires ${\sim}q \times 10$ relaxation steps for its internal Markov chain (where $q$ is the number of Potts states), yielding a total energy of ${\sim}0.084$~nJ per Gibbs sweep for $L = 1{,}273$ Potts nodes.
Over 10 sweeps: $10 \times 0.084 \approx 0.84$~nJ.

We note that Potts (p-dit) hardware does not yet exist; only the Ising (p-bit) TSU is the near-term target.
For an example of how a Potts model could be implemented in CMOS hardware, see Whitehead et al.~\cite{whitehead2023}.

\subsection*{GPU FLOP counting}

\Cref{tab:flops} summarizes the FLOP estimates for each algorithm.
These are order-of-magnitude estimates; the key observation is that the algorithms span several orders of magnitude in computational cost for comparable solution quality.

\begin{table}[H]
\centering
\caption{FLOP estimates for each optimization algorithm on the spike protein ($L = 1{,}273$).}
\label{tab:flops}
\vspace{2mm}
\begin{tabular}{llrr}
\toprule
Algorithm & Per-step cost & Steps & Total FLOPs \\
\midrule
Potts (10 sweeps) & ${\sim}100$ FLOPs/position & $10 \times 1{,}273$ & ${\sim}1.3 \times 10^6$ \\
Ising ($2 \times 10^4$ sweeps) & ${\sim}32$ FLOPs/spin & $2 \times 10^4 \times 3{,}147$ & ${\sim}2.0 \times 10^9$ \\
GA (tuned) & see below & $200 \times 1{,}000$ & ${\sim}6 \times 10^9$ \\
\bottomrule
\end{tabular}
\end{table}

\paragraph{Potts model (${\sim}1.3 \times 10^6$ FLOPs).}
Each of the 10 Gibbs sweeps updates all $L = 1{,}273$ positions.
Per position, the computation involves: bias lookup ($K = 6$ values), two neighbor interaction lookups (indexing into a $K \times K$ pairwise matrix), softmax normalization over $K$ states (${\sim}3K$ operations for exponentiation and normalization), and a categorical sample (${\sim}K$ operations for cumulative sum and comparison).
This totals ${\sim}100$ FLOPs per position.

\paragraph{Ising model (${\sim}2.0 \times 10^9$ FLOPs).}
Each of the $2 \times 10^4$ Gibbs sweeps updates all $N = 3{,}147$ spins.
Per spin: computing the effective field from ${\sim}12$ neighbors requires ${\sim}24$ multiply-add operations, plus the bias term (1 operation), a sigmoid evaluation (${\sim}5$ operations for the approximation), and a Bernoulli sample (2 operations).
This totals ${\sim}32$ FLOPs per spin.

\paragraph{Genetic algorithm (${\sim}6 \times 10^9$ FLOPs).}
With population size 200 and $1{,}000$ generations, each generation involves:
\begin{itemize}
  \item \textit{Scoring} (${\sim}6L$ FLOPs per individual): GC counting (sum over $L$ positions), repeat penalty lookup ($L{-}1$ table lookups and additions), and rarity score lookup ($L$ table lookups).
  \item \textit{Crossover} (${\sim}5L$ FLOPs per offspring): parent selection (random index), uniform crossover mask generation and application.
  \item \textit{Mutation} (${\sim}15L$ FLOPs per offspring): for each position, generate a random number, compare against mutation rate, and when mutating, perform frequency-weighted codon sampling (cumulative probability lookup).
  \item \textit{Selection and sorting}: ranking the population by score.
\end{itemize}
The GA requires ${\sim}5{,}000\times$ more FLOPs than the Potts sampler for comparable solution quality.

\subsection*{GPU energy bounds}

We estimate GPU energy on an NVIDIA RTX~3090 (350~W TDP, 35.6~TFLOPS FP32 peak) using two methods:

\paragraph{Idealized lower bound.}
The idealized energy is computed as:
\begin{equation}
  E_{\mathrm{ideal}} = \frac{\text{Total FLOPs}}{\text{Peak FLOP/s}} \times \text{TDP}
  = \frac{\text{Total FLOPs}}{35.6 \times 10^{12}} \times 350~\text{W}
\end{equation}
This is a generous lower bound: no real workload sustains 100\% utilization, and the codon optimization workloads are far too small to saturate a modern GPU.

\paragraph{Measured upper bound.}
The measured energy multiplies the wall-clock execution time of our JIT-compiled JAX implementation by TDP:
\begin{equation}
  E_{\mathrm{measured}} = t_{\mathrm{wall}} \times 350~\text{W}
\end{equation}
This is an upper bound since the GPU may not draw full TDP power during execution.

\Cref{tab:gpu-energy} summarizes both estimates.
The large gap between idealized and measured energy (2--4 orders of magnitude for the Potts model) reflects the fact that these workloads are too small to efficiently utilize GPU hardware: execution time is dominated by kernel launch overhead, memory latency, and pipeline stalls rather than by floating-point computation.

\begin{table}[H]
\centering
\caption{GPU energy estimates on an NVIDIA RTX~3090 (350~W TDP, 35.6~TFLOPS FP32 peak). The Ising wall time was estimated by scaling from an RTX 4080 Laptop measurement.}
\label{tab:gpu-energy}
\vspace{2mm}
\begin{tabular}{lrrrr}
\toprule
Algorithm & FLOPs & Idealized energy & Wall time & Measured energy \\
\midrule
Potts (10 sweeps) & $1.3 \times 10^6$ & 13~\textmu J & 0.7~ms & 245~mJ \\
Ising ($2 \times 10^4$ sweeps) & $2.0 \times 10^9$ & 20~mJ & ${\sim}240$~ms & ${\sim}84$~J \\
GA (tuned) & $6 \times 10^9$ & 58~mJ & 60~ms & 21~J \\
\bottomrule
\end{tabular}
\end{table}

\section*{Supplementary Note 3: Algorithm details and hyperparameters}

This note provides full hyperparameter tables and algorithmic details for reproducibility.

\subsection*{Hyperparameter tables}

\Cref{tab:potts-params,tab:ising-params,tab:ga-params} list all parameters used for the results reported in the main text.

\begin{table}[H]
\centering
\caption{Potts model hyperparameters.}
\label{tab:potts-params}
\vspace{2mm}
\begin{tabular}{ll}
\toprule
Parameter & Value \\
\midrule
Sequence & SARS-CoV-2 spike protein ($L = 1{,}273$) \\
Host organism & \emph{E. coli} K-12 \\
Codon usage weight $w_f$ & 0.1 \\
GC content weight $w_{\mathrm{GC}}$ & 1.0 (standard) / $2 \times 10^4$ (hard) \\
Repeat penalty weight $w_R$ & 0.1 (standard) / 0.2 (hard) \\
Target GC fraction $\rho_T$ & 0.5 \\
Number of Potts states $K_{\mathrm{max}}$ & 6 (padded) \\
Annealing steps $T$ & 10 \\
$\beta_{\mathrm{min}}$ & 0.3 \\
$\beta_{\mathrm{max}}$ & 100 \\
Gibbs sweeps per step & 1 \\
Total Gibbs sweeps & 10 \\
Adaptive GC multiplier $\eta$ & 1.0 \\
Block decomposition & 2-color (even/odd positions) \\
Independent chains & 512 \\
\bottomrule
\end{tabular}
\end{table}

\begin{table}[H]
\centering
\caption{Ising model hyperparameters.}
\label{tab:ising-params}
\vspace{2mm}
\begin{tabular}{ll}
\toprule
Parameter & Value \\
\midrule
Energy function weights & Same as Potts (\cref{tab:potts-params}) \\
Total spins $N$ & 3{,}147 \\
Average spins per position & 2.47 \\
Maximum node degree & 12 \\
Annealing steps $T$ & 10 \\
Sweeps per step & 2{,}000 (standard) / 4{,}000 (hard) \\
Total Gibbs sweeps & $2 \times 10^4$ (standard) / $4 \times 10^4$ (hard) \\
$\beta_{\mathrm{min}}$, $\beta_{\mathrm{max}}$ & 2.0, 200 (standard) / 0.5, 150 (hard) \\
$P_{\mathrm{min}}$, $P_{\mathrm{max}}$ & 2.0, 200 (standard) / 4.0, 150 (hard) \\
Initial constant-$P$ steps $n_{\mathrm{const}}$ & 5 \\
Adaptive GC multiplier & 0.9 \\
Block decomposition & 4-color (position parity $\times$ spin-index parity) \\
Independent chains & 512 \\
\bottomrule
\end{tabular}
\end{table}

\begin{table}[H]
\centering
\caption{Genetic algorithm hyperparameters.}
\label{tab:ga-params}
\vspace{2mm}
\begin{tabular}{ll}
\toprule
Parameter & Value \\
\midrule
Population size & 200 \\
Generations & 1{,}000 \\
Elite survivors & 10 \\
Lucky survivors (random) & 2 \\
Mutation rate & 0.003 \\
Crossover type & Uniform \\
Initialization & Frequency-weighted random sampling \\
Independent chains & 512 \\
\bottomrule
\end{tabular}
\end{table}

\subsection*{GA parameter sensitivity}

The genetic algorithm's performance is highly sensitive to the mutation rate.
The default mutation rate of $0.05$ from Fox et al.~\cite{fox2021mrna} yields scores in the range ${\sim}550$--$600$ on the spike protein under the hard parameter setting, and comparably degraded performance under the standard weights.
Reducing the mutation rate to $0.003$ is required to achieve scores comparable to the Potts and Ising models under both parameter settings.

This sensitivity has a natural interpretation: the uniform crossover operator already provides substantial exploration by recombining partial solutions from different individuals.
A high mutation rate is counterproductive because it disrupts good partial solutions faster than selection can preserve them.
The optimal mutation rate balances the need to introduce novel codon choices (which crossover alone cannot produce) against the cost of destroying beneficial codon combinations that selection has accumulated.

We identified the optimal parameters through a systematic sweep over mutation rate ($0.001$--$0.1$), population size ($50$--$500$), and number of elite survivors ($5$--$20$), evaluating each configuration over multiple seeds.

\subsection*{Annealing schedules}

Both the Potts and Ising models use simulated annealing~\cite{kirkpatrick1983sa} with a log-spaced inverse temperature schedule:
\begin{equation}\label{eq:beta-schedule}
  \beta_t = \beta_{\mathrm{min}} \cdot \left( \frac{\beta_{\mathrm{max}}}{\beta_{\mathrm{min}}} \right)^{t/T}
\end{equation}
for $t \in \{0, \ldots, T\}$, where $T$ is the number of annealing steps.
At each step, the host updates $\beta$, recomputes the adaptive GC coefficient, and runs a fixed number of Gibbs sweeps.

For the Ising model, the constraint penalty $P$ follows an independent schedule: it stays constant at $P_{\mathrm{min}}$ for the first $n_{\mathrm{const}}$ steps, then ramps logarithmically:
\begin{equation}\label{eq:P-schedule}
  P_t = \begin{cases}
    P_{\mathrm{min}} & t \leq n_{\mathrm{const}} \\[4pt]
    P_{\mathrm{min}} \cdot \left( \dfrac{P_{\mathrm{max}}}{P_{\mathrm{min}}} \right)^{(t - n_{\mathrm{const}})/(T - n_{\mathrm{const}})} & t > n_{\mathrm{const}}
  \end{cases}
\end{equation}
Low $P$ early in annealing permits invalid-state (defect) transitions that provide shortcuts around energy barriers: when a valid intermediate state $k$ has energy exceeding $P$, the defect route (costing $P$) is energetically cheaper.
High $P$ at the end enforces strict constraint satisfaction.

\subsection*{Spike protein spin count}

The spike protein ($L = 1{,}273$) maps to $N = 3{,}147$ domain-wall spins, an average of $2.47$ spins per amino acid position.
This is less than the maximum of $K_{\mathrm{max}} - 1 = 5$ because many amino acids have fewer than 6 synonymous codons (e.g., methionine and tryptophan have only 1 codon each and contribute 0 spins).

\section*{Supplementary Note 4: Adaptive GC coefficient}\label{sec:supp_gc}

This note gives the full derivation of the adaptive GC coefficient scheme summarised in Section~4.3 of the main text.
The scheme is applied identically in the Potts and Ising formulations; the only differences are (i)~in the Ising case the linear GC bias is compiled through the domain-wall encoding into Ising biases (Supplementary Note~1), and (ii)~the Ising case relies on the GC-sorted codon ordering described at the end of this note.

\subsection*{Why the GC term cannot be implemented natively}

The GC content term from the energy function (Eq.~(4) of the main text) couples all positions through a single global quadratic constraint:
\begin{equation}\label{eq:gc-term-supp}
  E_{\mathrm{GC}}(\mathbf c) = w_{\mathrm{GC}} \Big( \tfrac{1}{3L} \textstyle\sum_{p=1}^{L} g(c_p) - \rho_T \Big)^{\!2}.
\end{equation}
Expanding the square yields an \emph{all-to-all} pairwise interaction between the GC counts of every pair of positions:
\begin{equation}\label{eq:gc-expanded}
  E_{\mathrm{GC}} = \frac{w_{\mathrm{GC}}}{(3L)^2} \sum_{p,q} g(c_p)\, g(c_q)
  \;-\; \frac{2 w_{\mathrm{GC}} \rho_T}{3L} \sum_p g(c_p)
  \;+\; \mathrm{const}.
\end{equation}
Representing this interaction directly on a TSU would require $\binom{L}{2}$ dense couplings.
Two fatal problems follow:
\begin{itemize}
  \item \textbf{Connectivity.}
    The near-term TSU target supports only sparse, locally-connected graphs (see Section~4.7 of the main text).
    Embedding a complete graph on $L = 1{,}273$ nodes into that fabric is not feasible.
  \item \textbf{Chromatic number.}
    Even if the dense connectivity were available, a complete graph on $L$ nodes has chromatic number $L$.
    Block Gibbs sampling requires as many sequential update phases per sweep as the chromatic number of the interaction graph, so a dense GC term would collapse the parallel block Gibbs scheme to fully sequential updates---destroying the entire advantage of the thermodynamic chip.
\end{itemize}
We therefore replace the global quadratic term with a \emph{per-chain linear} approximation that is refreshed between annealing steps.
This keeps the graph sparse and 2-colorable (Potts chain) or 4-colorable (Ising DWE), while still steering each chain toward the target GC fraction $\rho_T$.

\subsection*{Linear approximation and update rule}

Let $\rho(\mathbf c) \mathrel{:=} \frac{1}{3L} \sum_p g(c_p)$ denote the GC fraction of a configuration, and let $\lambda$ be a scalar coefficient, maintained \emph{independently per Markov chain}.
We fold $\lambda$ into the Potts unary biases as
\begin{equation}\label{eq:lambda-bias}
  \tilde h_p(k) \;=\; h_p(k) \;+\; \lambda \, g\!\left(c_p^{(k)}\right),
\end{equation}
where $h_p(k)$ is the original Potts bias (codon-usage rarity plus, in the Ising case, the DWE constraint contribution).
Positive $\lambda$ favours GC-rich codons, negative $\lambda$ penalises them.

At each annealing step $t$ the host reads the current GC fraction $\rho_t$ of each chain (computed in a few operations from the last observed state) and updates $\lambda$ via
\begin{equation}\label{eq:lambda-update}
  \lambda_{t+1} \;=\; \lambda_t \;-\; \eta\,(\rho_t - \rho_T),
\end{equation}
where $\eta > 0$ is an adaptation rate (\texttt{gc\_coeff\_adapt\_mult} in the code).
If the chain has overshot the target ($\rho_t > \rho_T$), $\lambda$ is pushed negative so that the next sampling burst penalises further GC incorporation; if it has undershot, $\lambda$ is pushed positive.
The update is entirely host-side between TSU flashes.

\subsection*{Clamping to the quadratic gradient}

\Cref{eq:lambda-update} on its own is unstable over a long run: $\lambda$ can drift well past any value that would actually be implied by the quadratic penalty, at which point the linear GC bias dominates the codon-usage and repeat terms and produces pathological solutions.
We therefore bound $\lambda_{t+1}$ at each step by the gradient of the true quadratic GC term, evaluated at the current GC fraction.
From \cref{eq:gc-term-supp},
\begin{equation}\label{eq:lambda-true}
  \lambda_{\mathrm{true}}(\rho_t)
  \;\mathrel{:=}\; -\,\frac{\partial E_{\mathrm{GC}}}{\partial (\text{GC count})}\bigg|_{\rho_t}
  \;=\; -\,\frac{2 w_{\mathrm{GC}} \,(\rho_t - \rho_T)}{3L}.
\end{equation}
This is the coefficient of the linearisation of \cref{eq:gc-term-supp} around $\rho_t$: adding $\lambda_{\mathrm{true}} \cdot g(c_p^{(k)})$ to every Potts bias reproduces the first-order effect of the true quadratic penalty on a codon's relative weight at $\rho_t$.
Crucially, the sign of $\lambda_{\mathrm{true}}$ is always aligned with the sign of the increment in \cref{eq:lambda-update}: both push $\lambda$ negative when $\rho_t > \rho_T$ and positive when $\rho_t < \rho_T$.
We define
\begin{equation}
  \lambda_{t+1}^{\,\mathrm{unclamped}} \;\mathrel{:=}\; \lambda_t - \eta\,(\rho_t - \rho_T),
\end{equation}
and then clip to the interval between $\lambda_t$ and $\lambda_{\mathrm{true}}(\rho_t)$:
\begin{equation}\label{eq:clip}
\boxed{\;
  \lambda_{t+1} \;=\; \mathrm{clip}\!\left(\lambda_{t+1}^{\,\mathrm{unclamped}},\;
    \min\!\big(\lambda_t,\,\lambda_{\mathrm{true}}(\rho_t)\big),\;
    \max\!\big(\lambda_t,\,\lambda_{\mathrm{true}}(\rho_t)\big)\right).
\;}
\end{equation}
This enforces two simultaneous constraints:
\begin{itemize}
  \item \textbf{Bound by $\lambda_{\mathrm{true}}$ (do not overshoot).}
    $\lambda_{t+1}$ never passes $\lambda_{\mathrm{true}}(\rho_t)$.
    So the linear surrogate is never stronger, at the current operating point, than the quadratic term it is replacing.
  \item \textbf{Monotonic progression (do not back off).}
    $\lambda_{t+1}$ also never retreats past $\lambda_t$.
    This preserves the accumulated pressure on the chain: if a string of previous steps has driven $\lambda$ strongly negative, a single noisy dip in $\rho_t$ does not immediately undo the correction.
\end{itemize}
The clamp therefore forces $\lambda$ to move monotonically toward $\lambda_{\mathrm{true}}(\rho_t)$ from whichever side it starts on.
The update is implemented in \texttt{adapt\_gc\_coeff} (\path{codons/problem.py}).

The per-chain nature of $\lambda$ is essential: early in annealing, different chains land in very different basins and thus have different $\rho_t$.
A single shared $\lambda$ would push all chains with the same force regardless of where they actually are, defeating the purpose of running many chains in parallel.
In our experiments each of the 512 chains therefore carries its own $\lambda$, updated independently via \cref{eq:clip}.

\subsection*{Why not just set $\lambda = \lambda_{\mathrm{true}}$?}\label{sec:gc-osc}

The derivation of $\lambda_{\mathrm{true}}(\rho_t)$ in \cref{eq:lambda-true} is tempting to apply directly: at each step, read off $\rho_t$, set $\lambda_{t+1} = \lambda_{\mathrm{true}}(\rho_t)$, and let the sampler do the rest.
This would replace the integrator of \cref{eq:lambda-update,eq:clip} with an instantaneous ``re-fit'' of $\lambda$ to the current GC fraction, and at first glance it should converge at least as fast.
It does not. The direct rule is unstable whenever the GC term is strong enough to materially constrain the optimisation, and in that regime collapses into a period-2 limit cycle in which $\rho$ and $\lambda$ bounce between two widely separated values on every single annealing step.

\begin{figure}[H]
\centering
\includegraphics[width=0.95\textwidth]{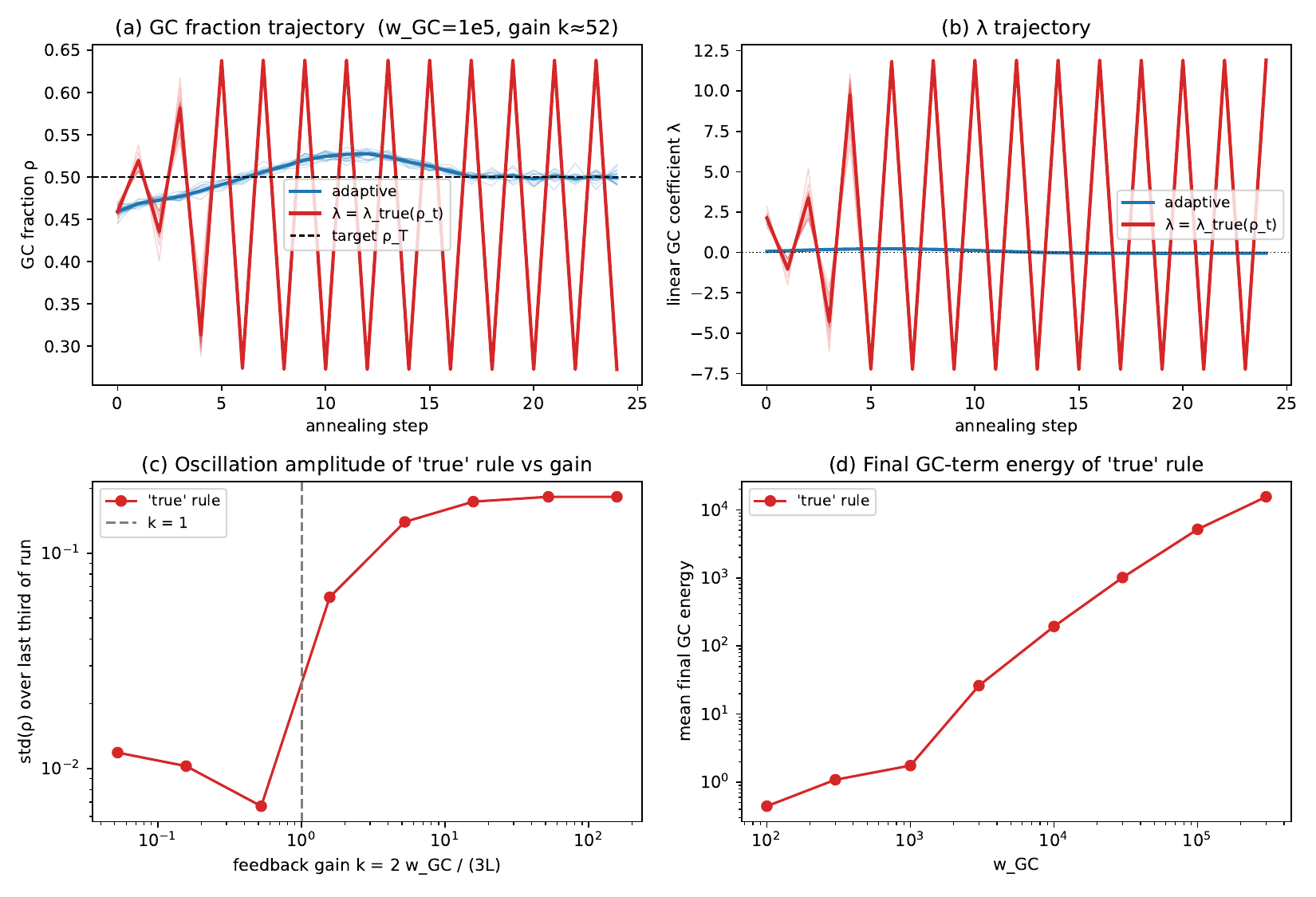}
\caption{\textbf{The direct rule $\lambda = \lambda_{\mathrm{true}}(\rho_t)$ is unstable.}
\textbf{(a,b)}~GC fraction $\rho$ and linear coefficient $\lambda$ across annealing steps on the spike protein, $w_{\mathrm{GC}} = 10^5$ (hard setting).
Light traces: 16 individual chains. Bold: mean over chains.
The adaptive rule (blue) quickly settles to $\rho \approx \rho_T = 0.5$ with $\lambda \approx {-}0.05$; the direct rule (red) locks into a period-2 limit cycle with $\rho$ alternating between $0.27$ and $0.64$ and $\lambda$ alternating between $+11.9$ and $-7.3$ every single step.
\textbf{(c,d)}~Sweep of $w_{\mathrm{GC}}$ for the direct rule. Oscillation amplitude (std of $\rho$ across the last third of the run) is noise-level below the predicted threshold $k = 2 w_{\mathrm{GC}} / (3L) = 1$ (dashed line) and saturates at the sampler's physical response range ($\approx 0.18$) well above it. The final GC-term energy grows by orders of magnitude across the same sweep.}
\label{fig:gc-osc}
\end{figure}

\paragraph{Empirical failure mode.}
\Cref{fig:gc-osc} shows representative trajectories under the hard weight setting ($w_{\mathrm{GC}} = 10^5$, $w_f = 0.1$, $w_R = 0.2$) on the spike protein, $L = 1{,}273$, 32 chains, 25 annealing steps with 15 Gibbs sweeps per step.
Under the direct rule, after a short transient of only ${\sim}5$ steps every chain enters an exact period-2 cycle:
$\rho_t \in \{0.273,\,0.638\}$ alternates on successive steps, and the corresponding $\lambda_t \in \{+11.91,\,{-}7.25\}$.
Those two $\lambda$ values are exactly $\lambda_{\mathrm{true}}(0.273)$ and $\lambda_{\mathrm{true}}(0.638)$, so the cycle is self-consistent: each step the rule prescribes the $\lambda$ that precisely drives the sampler to the \emph{other} state of the cycle.
The sign of $\rho_{t+1}-\rho_T$ flips at every single step (``flip rate'' = $1.00$), the mean absolute deviation $|\rho - \rho_T|$ over the last third of the run is ${\approx}0.19$ (vs $0.005$ for the adaptive rule), and the mean total energy over the 2-cycle is ${\sim}4{,}000$ (vs ${\sim}450$ for the adaptive rule): optimisation fails outright.

\paragraph{Linear stability analysis.}
The direct rule is a pure proportional feedback controller:
\begin{equation}\label{eq:direct-rule}
  \lambda_{t+1} \;=\; -k\,(\rho_t - \rho_T),
  \qquad
  k \;\mathrel{:=}\; \frac{2\,w_{\mathrm{GC}}}{3L}.
\end{equation}
The loop is closed by the sampler: given enough Gibbs sweeps at inverse temperature $\beta_t$, the new bias $\lambda_{t+1}$ drives the chain to a new equilibrium GC fraction $\rho_{t+1}$.
For a single categorical position $c_p$ whose Gibbs distribution is $\propto \exp(\beta(h_p(k) + \lambda\, g(c_p^{(k)})))$ (THRML's weight convention), the standard exponential-family identity gives
$\partial\langle g(c_p)\rangle/\partial\lambda = \beta \,\mathrm{Var}_p(g)$, and summing over positions yields the per-chain susceptibility
\begin{equation}\label{eq:susceptibility}
  \frac{\partial \rho}{\partial \lambda}\bigg|_{\lambda=0}
  \;=\; \frac{\beta_t}{3L}\sum_{p=1}^{L} \mathrm{Var}_p\!\big(g(c_p)\big)
  \;=:\; \chi(\beta_t).
\end{equation}
Linearising \cref{eq:direct-rule} together with the sampler response around $\rho_T$ gives
\begin{equation}\label{eq:linear-loop}
  \rho_{t+1} - \rho_T \;\approx\; -A(\beta_t)\,(\rho_t - \rho_T),
  \qquad
  A(\beta_t) \;=\; k\cdot \chi(\beta_t).
\end{equation}
The fixed point $\rho = \rho_T$ is stable iff $|A| < 1$ and unstable otherwise; the sign of $A$ is always positive, so as soon as $A > 1$ the iteration produces a growing two-sided overshoot, i.e.\ a geometrically expanding 2-cycle.

For the codon problem on \emph{E.\ coli}-weighted tables, the per-position variance of the GC count $\mathrm{Var}_p(g)$ averages to $\mathcal O(0.3)$ at moderate $\beta_t$, giving $\chi(\beta_t) \sim 0.1\,\beta_t$ as an order-of-magnitude estimate (the true $\chi$ grows sublinearly at large $\beta_t$ because the per-position distribution concentrates on the locally-favoured codon and its variance decreases).
To leading order,
\begin{equation}\label{eq:A-estimate}
  A(\beta_t) \;\sim\; 0.1\,\beta_t \cdot \frac{2 w_{\mathrm{GC}}}{3L}.
\end{equation}
The annealing schedule spans $\beta_t \in [0.3, 100]$.
At the hard-weight setting $w_{\mathrm{GC}} = 10^5$ ($k \approx 52$), this estimate gives $A$ in the range $\approx 1.6$ to $\approx 520$ -- unstable from the very first step.
At the standard Fox weights $w_{\mathrm{GC}} = 1$ ($k \approx 5\!\times\!10^{-4}$), $A$ never exceeds $\approx 5\!\times\!10^{-3}$, and the direct rule would in fact work there.
The oscillation only arises in regimes where the GC term is heavily enforced -- which is precisely the regime in which any GC correction is needed at all, and in particular the hard-weight setting used for the stress tests in the main text.

\paragraph{Threshold verified by sweep.}
\Cref{fig:gc-osc}c,d sweeps $w_{\mathrm{GC}}$ through two-and-a-half orders of magnitude for the direct rule.
The transition between the stable and oscillating regimes occurs sharply at $k \approx 1$ (between $w_{\mathrm{GC}} = 10^3$ and $3\!\times\!10^3$), as predicted by \cref{eq:linear-loop}. Above that threshold the amplitude grows until it saturates at $\approx 0.18$: with a bias of magnitude $|\beta \lambda| \gg 1$, the sampler pushes every position to its lowest- or highest-GC codon, and the feasible range of $\rho$ over the codon table is the physical ceiling on the 2-cycle's amplitude.

\paragraph{Why the adaptive rule is stable.}
Replacing \cref{eq:direct-rule} with the integral rule of \cref{eq:lambda-update} gives a two-dimensional linear map in $(\rho - \rho_T,\,\lambda)$ whose Jacobian has eigenvalues $\{0,\;1 - \eta\,\chi(\beta_t)\}$.
The fixed point is stable whenever $\eta\,\chi(\beta_t) \in (0, 2)$.
Crucially, this closed-loop gain is $\eta\,\chi$, \emph{independent of} $w_{\mathrm{GC}}$: the integrator decouples the per-step feedback gain from the strength of the quadratic penalty it is emulating.
Since $\chi(\beta_t)$ is an $\mathcal O(1)$ quantity bounded over the entire annealing schedule -- at small $\beta_t$ it scales linearly with slope ${\sim}0.1$ (the average codon-table GC variance divided by three), at large $\beta_t$ it decays back toward $0$ as each position's distribution concentrates on the locally-favoured codon -- a choice of $\eta = \mathcal O(1)$ keeps $\eta\chi$ safely inside $(0, 2)$ for \emph{any} $w_{\mathrm{GC}}$. Our production values ($\eta = 1.0$ for Potts and $\eta = 0.9$ for Ising; \cref{tab:potts-params,tab:ising-params}) sit comfortably in this range.

The clamp of \cref{eq:clip} plays a secondary but important role: the integrator, left alone, could accumulate $\lambda$ past $\lambda_{\mathrm{true}}$ at large $\beta_t$.
At that point the linear surrogate would be \emph{stronger} than the quadratic term it is replacing, and its competition with the codon-usage and repeat terms would distort the objective.
The clamp to $\lambda_{\mathrm{true}}(\rho_t)$ is the natural upper bound: once $\lambda$ reaches $\lambda_{\mathrm{true}}$, the linear approximation is already as strong as the true quadratic gradient at the current $\rho_t$, and further integration would over-penalise GC deviation.

In short, the right way to use $\lambda_{\mathrm{true}}$ is as a \emph{bound}, not as a \emph{target}: the adaptive rule walks monotonically toward $\lambda_{\mathrm{true}}$ from the previous value via small integrator steps, while the direct rule jumps to $\lambda_{\mathrm{true}}$ every step -- and, because the sampler's response gain $\chi$ is large enough to make $k\chi > 1$ at interesting weights, those jumps do not damp out but lock into a 2-cycle instead.

\subsection*{Integration with annealing and flash-sample-read}

Because $\lambda$ enters the Potts bias linearly, it is scaled by the current inverse temperature $\beta_t$ before being compiled into the model. The model biases at step~$t$ become
\begin{equation}\label{eq:compiled-bias}
  \tilde h_p^{(t)}(k) \;=\; \beta_t \,\Big(\, h_p(k) + \lambda_t \, g\!\left(c_p^{(k)}\right)\,\Big).
\end{equation}
In the Ising case, $\tilde h_p^{(t)}$ is subsequently compiled to Ising biases via domain-wall encoding (Supplementary Note~1, \cref{eq:bias,eq:bias-pairwise}).
The host algorithm at each annealing step is therefore:
\begin{enumerate}
  \item Read back the state of every chain from the TSU (or its simulator) and compute $\rho_t$ per chain.
  \item Update $\lambda_t \to \lambda_{t+1}$ via \cref{eq:clip}.
  \item Combine the new $\lambda_{t+1}$ with the scheduled $\beta_{t+1}$ (and, in the Ising case, $P_{t+1}$) to produce the biases of \cref{eq:compiled-bias}.
  \item Flash the new weights to the TSU and run \texttt{steps\_per\_beta} Gibbs sweeps.
\end{enumerate}
The cost on a real TSU is \emph{zero} additional flashes: weights would have to be re-flashed every annealing step anyway in order to update $\beta_t$ (and $P_t$ in the Ising case), so folding the updated $\lambda_{t+1}$ into the same flash adds no new operations.
The additional host work consists only of one reduction per chain (to compute $\rho_t$) and a handful of scalar operations.

\subsection*{GC-sorted codon ordering}

In the Ising / domain-wall formulation, a single spin flip moves the underlying Potts variable between two \emph{adjacent} indices, $k \leftrightarrow k{\pm}1$ (see Section~4.4 of the main text).
For the linear GC bias in \cref{eq:lambda-bias} to actually guide the domain wall, the codon indices at each position must be arranged so that adjacent indices differ by as little as possible in GC count.
Otherwise, moving the domain wall by one spin would cause discontinuous jumps in GC content, which the linear approximation cannot compensate for smoothly.

We therefore order the codons at each position by ascending GC count, with alphabetical tiebreaking among codons of equal GC count.
This is the ordering returned by \texttt{SORTED\_CODON\_TABLE} in \path{codons/problem.py}, and it is used in both the Potts and Ising formulations (the Potts chain does not require it, but using the same ordering keeps weight computation consistent between the two models).
\Cref{tab:codon-ordering} shows the resulting ordering for three representative amino acids.

\begin{table}[H]
\centering
\caption{GC-sorted codon ordering for representative amino acids. Index 0 is the lowest-GC codon; adjacent indices differ by at most 1 in GC count. The number in parentheses is the GC count of each codon.}
\label{tab:codon-ordering}
\vspace{2mm}
\begin{tabular}{llccccccl}
\toprule
Amino acid & \#Codons & Idx 0 & Idx 1 & Idx 2 & Idx 3 & Idx 4 & Idx 5 & DW spins \\
\midrule
Leucine (L) & 6 & TTA (0) & CTA (1) & CTT (1) & TTG (1) & CTC (2) & CTG (2) & 5 \\
Alanine (A) & 4 & GCA (2) & GCT (2) & GCC (3) & GCG (3) & -- & -- & 3 \\
Phenylalanine (F) & 2 & TTT (0) & TTC (1) & -- & -- & -- & -- & 1 \\
\bottomrule
\end{tabular}
\end{table}

\noindent
Under this ordering every single spin flip changes the GC content of the sequence by at most 1, and flips between codons of equal GC count are GC-neutral.
Two consequences follow.
First, the GC landscape seen by the domain wall becomes locally smooth: the term $\lambda \cdot g(c_p^{(k)})$ is monotone (up to ties) in the Potts index $k$, so after conversion to Ising biases via first differences (\cref{eq:bias}) the linear GC coefficient produces a spatially uniform shift along each spin chain rather than a sign-alternating pattern.
Second, the adaptive scheme of \cref{eq:clip} acquires a natural interpretation in spin space: $\lambda$ is a single scalar knob that uniformly shifts the bias along every domain wall, continuously trading off the cost of moving the wall one step inward against the benefit (or cost) of gaining one unit of GC.
An arbitrary codon ordering would scramble this relationship and require a per-codon schedule to achieve the same effect, which is incompatible with a single global $\lambda$.

\bibliography{refs}